\documentclass[a4paper,12pt]{JHEP}

\usepackage{epsfig}

\usepackage{fancyheadings}
\usepackage{cite}
\usepackage{amsmath}
\usepackage{amssymb}

\usepackage[latin1]{inputenc}
\usepackage[T1]{fontenc}
\usepackage{slashed}

\def\be{\begin{equation}}
\def\ee{\end{equation}}

\def\t{\widetilde}

\renewcommand{\d}{\mathrm{d}}
\newcommand{\e}{\mathrm{e}}
\newcommand{\co}{\cosh{\mu y}}
\newcommand{\si}{\sinh{\mu y}}

\newcommand{\R}{\mathcal{R}}
\newcommand{\C}{\mathcal{C}}
\newcommand{\HU}{\mathcal{H}}

\title{Exact Evolution of the Complete Non-Linear Cosmological Radion}

\author{
Ian R. Vernon\footnote{I.R.Vernon@dur.ac.uk},
Anne-Christine Davis\footnote{A.C.Davis@damtp.cam.ac.uk}, \\
\normalsize \em Centre for Particle Theory, University of Durham, \\
\normalsize \em South Road, Durham, DH1 3LE, U.K. \\
\normalsize \em Department of Applied Mathematics and Theoretical Physics,\\
\normalsize \em CMS, University of Cambridge, Cambridge, CB3 0WA, UK. \\
}

\abstract{The cosmological two brane model is discussed in detail. An elegant method for 
deriving the generalised non-linear equations for the interbrane distance is presented.
These equations are used to identify equilibrium positions of the radion, and are then 
linearised in order to determine the nature of these equilibrium solutions. Numerical 
methods are then employed to determine the complete behaviour of the (unstabilised) 
cosmological radion for a variety of interesting cases, including radiation and 
matter dominated non-$Z_2$ symmetric branes. The implications to 
theories involving brane collisions are then discussed.}

\preprint{DCPT-04/01}


\begin{document}
%

\section{Introduction}
\label{radion}

Recently there has been considerable interest in the novel suggestion
that we live in a Universe that possesses more than four
dimensions. The standard model fields  
are assumed to be confined to a hyper-surface (or 3-brane)
embedded in this higher dimensional space, in contrast to the gravitational
fields which propagate through the whole of
spacetime~\cite{ruba,akam1,anton,visser,ahdd,ahddII,aahdd,sund1,rscompact99,rshierarchy99}.
In order for this to be a
phenomenologically relevant model of our universe, 
standard four-dimensional gravity must be recovered on our
brane. There are various ways to do this, the most obvious being to
assume that the extra dimensions transverse to our brane are compact.
In this case gravity can be recovered on scales larger than the size
of the extra dimensions~\cite{ahdd,ahddII,aahdd}. This is different
from earlier proposals since the restrictions on the size of the 
extra dimensions from particle
physics experiments no longer apply, as the standard model fields 
are confined to the brane. The extra dimensions only have to
be smaller than the scale on which gravity experiments have probed,
of order 0.1mm at the time of writing.
Another way to recover four-dimensional gravity at large distances is 
to embed a positive tension 3-brane into an
AdS$_5$ bulk~\cite{rscompact99,rshierarchy99}. In this scenario four-dimensional
gravity is obtained at scales larger than the AdS radius. Randall
and Sundrum showed that this could produce sensible gravity even if
the extra dimension was not compact.

Several aspects of the above extra dimensional scenarios have since 
been investigated, and compared with the standard four-dimensional case.
The cosmology of a 3-brane in a five-dimensional bulk was studied and 
its Friedman equation derived and shown to contain several extra 
terms~\cite{noncosmo,bulkcosmo,cosmo1extrad,expans1extrad,Kraus}. Pertubations 
to this homogenious case have been examined~\cite{pert1,pert2,pert3,pert4} as 
have some inflationary models~\cite{inflation,infl2}, as well as 
phase transitions, topological defects and baryogenesis~\cite{cosmophase}. 
More recently it was shown how to embed the Randall-Sundrum models within 
supergravities~\cite{s49,s50,s51} and then within string theory compactifications 
as in~\cite{s52,s53,s54,s55}. The seemingly arbitrary feature of having an $AdS$ bulk 
spacetime is actually well motivated as it is found as a supersymmetric vacuum to 
supergravity theories, inspiring several more recent brane world 
models~\cite{anne1,anne2,anne3,anne4,anne5,anne6,anne7}. Generalisations have been made 
to six dimensions, and it has been 
demonstrated that both singular and non-singular self-gravitating string defects can 
localise gravity~\cite{greg1,shap6D,itsaso,s17,ruth111}.

In this paper we examine the `cosmological radion' which 
is defined as the interbrane distance of the cosmologically generalised 
Randall-Sundrum two brane model. The radion 
is in general time dependent and in this paper we go on to investigate this time 
dependent behaviour fully. Many authors impose a stabilisation mechanism 
on the radion such as~\cite{s8,s60,s59,stabmodulus,s62}, 
in order to have a static size of extra dimension.
While this is understandable, 
most of the mechanisms are inserted in an artificial manner and it is 
therefore interesting to examine the nature of the radion without such 
purpose built constraints. Several linearised approaches have been 
made~\cite{radionwave,stabmodulus,radionds,locallylocal,CsaRadStab} 
however, we will define the radion in a non-perturbative way, 
and then use the assumptions of homogeneity and isotropy to 
derive the non-linear equations of motion of the 
cosmological radion, thus providing a more general and elegant proof 
to that found in~\cite{bin3}.

The radion equations will be used to identify scenarios where stable and 
unstable interbrane distances exist, and then the equations will be 
linearised to determine the nature of these equilibrium positions. We then 
go on to solve numerically for the trajectories of the radion in several 
cases of interest including when both branes are $dS$ or $AdS$, when the 
reference brane possesses a sensible cosmology, when both branes are either 
radiation or matter dominated, when phase transitions 
occur, and when the $Z_2$ symmetry is broken allowing scenarios with 
two positive tensions branes. Hence we significantly 
extend the results found by~\cite{bin3}, in which only fairly trivial 
situations were looked at. A rich variety of behaviour of the radion is 
found and in order to 
understand this we present a conceptual argument based on the perspective 
of a bulk observer, summarised in section~\ref{bulkstuff1}.

\section{Deriving the Complete Non-Linear Equations for the
Cosmological Radion}

In this section we first describe the generalised Randall-Sundrum two brane 
scenario before defining the cosmological radion in a non-perturbative manner. 
The complete non-linear equation for the cosmological radion or interbrane
distance $\R(t)$ is then derived for the general case. It is assumed that no
$Z_2$ symmetry exists across either brane and that the
bulk possess a non-zero Weyl tensor component and we therefore
generalise the derivation presented in~\cite{bin3,ver3}.
Alternative, mainly linearised approaches can be found 
in~\cite{radionwave,stabmodulus,radionds,locallylocal,CsaRadStab}.

\subsection{Cosmological Two Brane Scenario}

Since we are interested in 
cosmological solutions, we take a metric of the form:
\begin{equation}\label{metA}
  \d s^2 \;=\; -n^2(t,y) \d t^2 \:+\: a^2(t,y) \gamma_{ij} \d x^i \d x^j \:+\: \d y^2,
\end{equation}
where $y$ is the coordinate of the fifth dimension and we adopt a 
brane-based approach where the reference brane is the hyper-surface defined by
$y=0$. Here $\gamma_{ij}$ is a maximally symmetric 3-dimensional
metric with $k=-1,0,1$ parameterising the spatial curvature. 
The metric is found 
by solving the five-dimensional Einstein's equations which take the form,
\begin{equation}\label{Ein11}
G_{AB} \equiv R_{AB} - \frac{1}{2}R g_{AB} = \kappa^2 T_{AB},
\end{equation}
where $R_{AB}$ is the five-dimensional Ricci tensor, $R=g^{AB}R_{AB}$ the 
scalar curvature and where we define $\kappa^2 = 1/\t{M}_5^3$, 
$\t{M}_5$ being the 
fundamental (reduced) 5D Planck Mass.
The stress-energy-momentum 
tensor can be written as,
\begin{equation}\label{stress11}
 T^A{ }_B = T^A{ }_B|_\mathit{brane} + T^A{ }_B|_\mathit{bulk}.
\end{equation}
Again with cosmology in mind, we assume a homogeneous and isotropic 
geometry in both branes and this makes it possible to write the first term as:
\begin{equation}
 T^A{ }_B|_\mathit{brane} \;\equiv\; 
\delta(y)\: {\rm diag}(-\rho_0,P_0,P_0,P_0,0) \;\; + \;\; 
\delta(y - \R(t))\: {\rm diag}(-\rho_2,P_2,P_2,P_2,0),
\end{equation}
where $\rho_0$ and $P_0$ are the energy density and pressure of the reference brane, 
with similar definitions for the second brane.
Here we have defined the position of the second brane to be at $y = \R(t)$ which is 
in general time dependent. $\R(t)$ is the interbrane distance otherwise known as the 
cosmological radion, and will be investigated throughout this paper.
The second term in equation (\ref{stress11}), which describes the same 
negative bulk cosmological constant $\Lambda$ either side of the brane, is of
the form,
\begin{equation}\label{coscom}
 T^A{ }_B|_\mathit{bulk} = 
    \frac{1}{\kappa^2}\:{\rm diag}(-\Lambda,-\Lambda,-\Lambda,-\Lambda,-\Lambda).
\end{equation}
Substituting the metric given by equation (\ref{metA}) into the Einstein 
equations (\ref{Ein11}) allows one to find the metric coefficients $a(t,y)$ 
and $n(t,y)$ in the bulk~\cite{fpaper}, which are used throughout this paper. 
The Friedmann equation on 
the reference brane can also be found by using the Einstein equations combined with 
the Israel junction conditions which relate the jump in the extrinsic curvature tensor to
the energy-momentum-tensor on the brane~\cite{s22,s23,s26,s27}. This gives the well 
known (non-$Z_2$ symmetric) brane world Friedmann equation~\cite{Kraus,fpaper} 
(with $a_0 \equiv a(t,y=0)$),
\begin{equation}
H_0^2 \; \equiv \; \left( 
      \frac{\dot{a}_0}{a_0}
\right)^2 \;=\; 
\frac{\Lambda}{6}   \; + \; 
\frac{\kappa^4}{36}\rho_0^2 \; - \; \frac{k}{a_0^2} \; + \; 
\frac{\mathcal{C}}{a_0^4}\; + \; \frac{F^2}{ \rho_0^2 a_0^8},
\end{equation}
where $\C$ and $F$ are constants related to the Weyl tensor and the lack of 
$Z_2$ symmetry respectively. These constants are in fact related to the black hole 
masses on either side of the reference brane by 
$\mathcal{C} = ( \mathcal{C}_0 + \mathcal{C}_1 ) /2$ and 
$F = 3( \mathcal{C}_1 - \mathcal{C}_0 )/ 4 \kappa^2 $, where we have defined the 
black hole mass to the left of the reference brane $(y<0)$ as $\mathcal{C}_0$ and 
to the right $(0<y<\R)$ as $\mathcal{C}_1$. Below we will also use the mass of the 
black hole to the right of the second brane $(\R < y)$ defined as $\mathcal{C}_2$.

In order to avoid the redundancy of several physical 
constants it is desirable to define a mass scale $\mu$ and an energy density scale 
$\sigma$ such that,
\begin{equation}
\mu^2 \;\; = \;\; -\frac{\Lambda}{6} \: , \quad \quad \quad \sigma 
\;\;= \;\; \frac{6\mu}{\kappa^2}\:.
\end{equation}
We can then deal solely with the dimensionless energy densities ($\eta_0$ and $\eta_2$)
and pressures ($p_0$ and $p_2$) of the branes, defined by,
\begin{eqnarray}\label{etadim1}
\eta_0 \;\; &=& \;\; \frac{\rho_0}{\sigma} \: , \quad \quad \quad 
\eta_2 \;\;\; = \;\;\; \frac{\rho_2}{\sigma} \: , \\
p_0 \;\; &=& \;\; \frac{P_0}{\sigma}\: , \quad \quad \quad \label{pdim1}
p_2 \;\;\; = \;\;\; \frac{P_2}{\sigma}\:.
\end{eqnarray}
These dimensionless quantities will be used throughout the rest of this paper. 
We now go on to derive the non-linear equation for $\R(t)$.

\subsection{An Elegant Method for Finding the Equation for $\dot{\R}$}

Using the dimensionless quantities defined by equations (\ref{etadim1}) and (\ref{pdim1}), 
the Friedmann equation on the reference brane now looks like,
\begin{equation}\label{5H_0}
H_0^2(t) \;\;\;=\;\;\; 
                   \mu^2 \eta_0^2 \;-\; \mu^2 \;-\; \frac{k}{a_0^2} \;+\; 
                   \frac{\mathcal{C}_0 + \mathcal{C}_1}{2 a_0^4} \;+\;
                   \frac{(\mathcal{C}_1 - \mathcal{C}_0)^2}
                        {16\mu^2 \eta_0^2 a_0^8} \, ,
\end{equation}
where $\C_0$ and $\C_1$ are the masses of the Schwarzschild black hole 
either side of the brane. If the second brane follows a trajectory given by
$y=\R(t)$, then the induced four-dimensional metric on the second brane
is given by,
\begin{eqnarray}
\d s^2 \;\;\; &=& \;\;\; -\left[
                        n^2(t,\R(t))- \dot{\R}^2
                        \right]\d t^2
                 + a^2(t,\R(t))\d x^2 , \\
              &=& \;\;\; -\d \tau^2 + a_2^2(\tau) \d x^2 ,
\end{eqnarray}
where the dot represents the derivative with respect to the reference brane time $t$, 
the subscript 2 implies evaluation on the second brane at $y=\R$, and 
$\tau$ has been defined as the proper time as seen by an
observer on the second brane and is defined by,
\begin{equation}\label{tau5}
\d \tau \;\;\; = \;\;\; \sqrt{
                        n^2(t,\R(t))- \dot{\R}^2
                               } \; \; \d t.
\end{equation}
The expansion rate of 
the second brane as seen by an observer on our brane is simply,
\begin{equation}
H_2(t) \;\;\; = \;\;\; \frac{1}{a_2}\frac{\d a_2}{\d t} 
       \;\;\; = \;\;\; \left(
                       \frac{\dot{a}}{a} + \frac{a'}{a}\dot{\R}
                       \right)_2,
\end{equation}
where the dash represents the derivative with respect to $y$, 
and as seen by an observer on the second brane itself,
\begin{eqnarray}\label{5H_2}
\mathcal{H}_2(\tau) \;\;\; &=& \;\;\; \frac{1}{a_2}
\frac{\d a_2}{\d \tau} \;\;\; = \;\;\; H_2(t) \frac{\d t}{\d \tau} , \\
\Rightarrow \;\;\; 
\mathcal{H}_2(\tau) \;\;\; &=& \;\;\;
                  \left(     
                       \frac{\dot{a}}{a} + \frac{a'}{a}\dot{\R}     
                  \right)_2                  
                  \left( 
                       n^2- \dot{\R}^2
                  \right)_2^{-1/2} .  \label{H25}           
\end{eqnarray}
This equation for $\HU_2$ is important as it relates the expansion
rate of the second brane (with respect to its proper time) to $\R$ and 
$\dot{\R}$. It does not seem so useful at first, as calculating $\HU_2$
could be difficult. However, it is known that the brane world Friedmann
equation given by (\ref{5H_0}) is derived from a purely local analysis
and that should we have chosen the second brane to be stationary and
at $y=0$, we would have derived the equivalent Friedmann equation with
$\eta_0$ replaced by $\eta_2$. This means that $\HU_2(\tau)$ must have
the following form,
\begin{equation}\label{5H_3}
\HU_2^2(\tau) \;\;\; = \;\;\; 
                \mu^2 \eta_2^2 \;-\; \mu^2  \;-\; \frac{k}{a_2^2} \;+\; 
                \frac{\mathcal{C}_1 +\mathcal{C}_2 }{2 a_2^4} \;+\;
                \frac{(\mathcal{C}_2 -\mathcal{C}_1)^2}
                     {16 \mu^2 \eta_2^2 a_2^8},
\end{equation}
where we have defined as before $\eta_2 = \rho_2/\sigma$ and $\C_1$
and $\C_2$ are 
again the masses of the black hole to the left and the right of the
second brane. 
It should be understood that equation (\ref{5H_3}) 
ensures that the second brane 
evolves according to the junction conditions of the
extrinsic curvature tensor, just as equation (\ref{5H_0}) ensures
similar behaviour for our brane. 
At this point we use the following bulk identity which 
is obtained from the Einstein equations and derived in~\cite{bulkcosmo},
\begin{equation}\label{eq:identity1}
\left(\frac{\dot{a}}{na} \right)^2 \;\; = \;\;
         \frac{a^{\prime 2}}{a^2} \;-\; \mu^2 \;-\; \frac{k}{a^2} \;+\; 
                \frac{\C_i}{a^4}\, ,
\end{equation}
where as before $i=0,1$ or $2$ for the regions $y<0, 0<y<\R,$ or
$y>\R$ respectively. Combining this with equation (\ref{5H_3}) gives,
for the $0<y<\R$ case,
\begin{equation}\label{5H_4}
\HU_2^2 \;\;\; = \;\;\; 
     \left(\frac{\dot{a}}{na} \right)_2^2 - \frac{a_2^{\prime 2}}{a_2^2} 
   +   \left( 
              \mu \eta_2 + \frac{f_2}{ \eta_2 a_2^4} 
        \right)^2,
\end{equation}
where the usual notation $f_2 = (\C_2-\C_1) / 4 \mu$ has been used.
Substituting (\ref{5H_4}) into (\ref{H25}) and rearranging gives
the following equation for $\dot{\R}$,
\begin{equation}\label{5R2}
\frac{a'}{a} \; + \; \frac{\dot{a}}{n^2 a} \dot{\R} \;\; = \;\; 
     \left(
           \mu \eta_2 + \frac{f_2}{ \eta_2 a_2^4}
     \right) 
       \left(1 - \frac{\dot{\R}^2}{n^2} \right)^{1/2},
\end{equation}
where we have dropped the subscript `2' as from now on all the metric 
components are to be evaluated at $y=\R$ unless stated otherwise. This therefore 
generalises the result derived by~\cite{bin3}.
Equation (\ref{5R2}) can be algebraically solved to give the generalised first order
equation for $\dot{\R}$:
\begin{equation}\label{5R1}
\dot{\R} \;\;\; = \;\;\;
 n \left[
         - \frac{a' \dot{a}}{a^2 n} \pm \HU_2
        \left(
              \mu \eta_2 + \frac{f_2}{ \eta_2 a_2^4}
        \right)
   \right]
   \left[ 
         \frac{a^{\prime 2}}{a^2} + \HU_2^2
   \right] ^{-1}.
\end{equation}
One has to be careful when considering in which situations solutions exist 
and in some cases which sign should be chosen in equation (\ref{5R1}).
We need to consider {\it both} equations (\ref{5R2}) and (\ref{H25}), a 
fact that was neglected by~\cite{bin3} and led to them having to 
change the $\pm$ sign in equation (\ref{5R1}) by hand in order to enable 
their numerical computation. Equation (\ref{5R2}) tells us immediately that 
$-n<\dot{\R}<n$, which just means that the motion of the second brane relative 
to the first cannot be greater than the local speed of light. With this in 
mind the same equation now tells us that if $a'/a$ and 
$\mu \eta_2 + f_2 / \eta_2 a_2^4$ are of the same sign, then if, 
$|a'/a| > |\dot{a}/na|$ there will be two solutions, however if 
$|a'/a| < |\dot{a}/na|$ only one solution exists. On the other hand, if 
$a'/a$ and $\mu \eta_2 + f_2 / \eta_2 a_2^4$ are of the opposite sign, 
$|a'/a| > |\dot{a}/na|$ implies no solutions and $|a'/a| > |\dot{a}/na|$ 
implies only one. For the cases where there is one solution only, the 
overall choice of the $\pm$ in (\ref{5R1}) should be the sign of 
$\dot{a}(\mu \eta_2 + f_2 / \eta_2 a_2^4)$. Inspecting equation (\ref{H25}) 
we see a similar story: if $\HU_2$ and $\dot{a}/n a$ are the same sign, 
then if $|a'/a|$ is greater/less than $|\dot{a}/na|$ it implies one/two 
solutions. If $\HU_2$ and $\dot{a}/n a$ are of opposite signs then if 
$|a'/a|$ is greater/less than $|\dot{a}/na|$ it implies one/zero 
solutions. The overall sign for the single solutions in this case should be 
that of $\HU_2 a'/a$. Therefore these two equations compliment each other: 
when (\ref{5R2}) implies there are two solutions, equation (\ref{H25}) 
shows that there is in fact only one (which one depends on the sign of 
$\HU_2$). This was neglected by~\cite{bin3}, and will be used in generating 
the numerical solutions in section~\ref{num5}.

\subsection{Derivation of the $\ddot{\R}$ Equation}

In order to investigate equilibrium positions of the second brane it
is necessary to derive the equation for $\ddot{\R}$. Taking the time
derivative of equation (\ref{5R2}) and then adding to it $H_2$ times
equation (\ref{5R2}) gives,
\begin{eqnarray}\label{derRdd}
&& \frac{\d}{\d t} \left(
                      \frac{a'}{a} + \frac{\dot{a}}{a n^2}\dot{\R}
               \right) + 
\left(
      \frac{\dot{a}}{a} + \frac{a'}{a}\dot{\R}
\right)
\left(
      \frac{a'}{a} + \frac{\dot{a}}{a n^2}\dot{\R}
\right) 
\;\; = \;\;  \\ \nonumber
&& \;\;\; \frac{1}{2} \: g \left(1 - \frac{\dot{\R}^2}{n^2} \right)^{-1/2} 
 \frac{\d}{\d t}  \left(
                        -\frac{\dot{\R}^2}{n^2}
                  \right) \; + \;
\frac{\d g}{\d t} \left(1 - \frac{\dot{\R}^2}{n^2} \right)^{1/2} + \;
g \: H_2 \left(1 - \frac{\dot{\R}^2}{n^2} \right)^{1/2},
\end{eqnarray}
where the function $g(a,\eta_2)$ is given by 
$g = \mu \eta_2 + f_2/\eta_2 a^4$. Noting first that 
\begin{equation}
\frac{\d g}{\d t} \;\; = \;\; 
                 \mu \dot{\eta}_2 
               - \frac{f_2}{\eta_2 a^4} \frac{\dot{\eta}_2}{\eta_2}
               - 4 \frac{f_2}{\eta_2 a^4} \frac{1}{a} \frac{\d a}{\d
t}\, ,
\end{equation}
and the identities: $(1/a)\d a/\d t = H_2$ and 
$\dot{\eta_2} = -3H_2(\eta_2 + p_2)$ which is just the energy-momentum 
conservation equation for the second brane, one then finds that by 
multiplying (\ref{derRdd}) by $1-\dot{\R}^2/n^2$ and rearranging, one gets,
\begin{eqnarray} \nonumber
&& - \; \frac{1}{3} G_{05}
               \left(
                     1 - \frac{\dot{\R}^4}{n^4}
               \right)
\; - \; \frac{1}{3} \dot{\R} 
              \left(
                    G_{55} + G_{00} / n^2
              \right)
               \left(
                     1 - \frac{\dot{\R}^2}{n^2}
               \right) \\ \nonumber && \; + \; 
H_2 \left[
          \frac{\ddot{\R}}{n^2} \;+\; \frac{n'}{n}
                     \left(
                           1-2\frac{\dot{\R}^2}{n^2}
                     \right)
           \;-\; \frac{\dot{n}}{n} \frac{\dot{\R}}{n^2}
    \right] \\ \label{derRdd2}
 \;\; = \;\;
&&  
 H_2  \left(
         -2\mu \eta_2 - 3\mu p_2 + 
          \frac{3 f_2 p_2}{ \eta_2^2 a^4}
   \right)
           \left(
                 1-\frac{\dot{\R}^2}{n^2}
           \right)^{3/2}.
\end{eqnarray}
We will be mainly interested in the case where the bulk possesses a 
cosmological constant, and therefore from the Einstein 
equations (\ref{Ein11}) we have 
$G_{05} = 0$ and $G_{00} / n^2 = - G_{55} = -6\mu^2$. 
This means that the first two terms 
on the left hand side of equation (\ref{derRdd2}) vanish. In fact a more 
general argument first proposed by~\cite{bin3} shows that these terms 
will vanish as long as the bulk energy-momentum tensor satisfies the 
three-dimensional symmetries of homogeneity and isotropy. 
>From (\ref{derRdd2}) we can now see that the generalised equation 
for $\ddot{\R}$ is given by,
\begin{equation}\label{5Rdd}
\frac{\ddot{\R}}{n^2} \;+\; \frac{n'}{n}
           \left(
                 1-2\frac{\dot{\R}^2}{n^2}
           \right)
 \;-\; \frac{\dot{n}}{n} \frac{\dot{\R}}{n^2} \;\; = \;\;
        \left(
              -2\mu \eta_2 - 3\mu p_2 + 
               \frac{3 f_2 p_2}{ \eta_2^2 a^4}
        \right)
           \left(
                 1-\frac{\dot{\R}^2}{n^2}
           \right)^{3/2}.
\end{equation} 
Equations (\ref{5R1}) and (\ref{5Rdd}) govern the evolution of $\R$ and
will be used to examine the behaviour of the general cosmological radion.

\section{The Linearised Radion Equation}\label{TLRE}

In this section the full equations (\ref{5R1}) and (\ref{5Rdd}) 
that govern the evolution of the cosmological radion are examined 
analytically. Unfortunately, due to the highly non-linear nature of
these equations finding an analytic solution is impossible in all but
the most trivial of cases. Therefore, here the stable solutions will be
identified and the nature of their stability discussed along with a
few other important features that the solutions must possess. These
results will then be compared with the extensive numerical analysis that
occurs in the next section.

\subsection{Equilibrium Solutions}

In order to find the equilibrium solutions, $\dot{\R}$ and $\ddot{\R}$
are both set to zero in equations (\ref{5R1}) and (\ref{5Rdd}). This
restricts the equation of state on the second brane to be of a
specific form. It will be shown in general that in order for the size
of the fifth dimension to be constant, an equation of state of the
form $p_2 = \omega_2 \eta_2$ inevitably leads to a time dependent
$\omega_2$.
Setting $\dot{\R} = \ddot{\R} = 0$ in equation (\ref{5R1}) leads to,
\begin{equation}\label{eta2conA}
\frac{a'}{a} \;\; = \;\; \mu \bar{\eta}_2 \; + \;
                         \frac{f_2}{\bar{\eta}_2 a^4},
\end{equation}
where $a$ and $a'$ are evaluated at position $y=\R$ and we have defined 
$\bar{\eta}_2$ to be the dimensionless energy density on the second brane 
at equilibrium. Rearranging gives,
\begin{equation}\label{eta2conB}
2\mu \bar{\eta}_2 \;\; = \;\; \frac{a'}{a} \; \pm \; 
                              \sqrt{
                                    \frac{a^{\prime 2}}{a^2}  
                                  - \frac{4 f_2 \mu}{a^4}
                                   }.
\end{equation}
This shows that if for example $a'/a < 0$ for all $y$ (as is the case
for a cosmologically realistic $Z_2$ symmetric brane where 
$a(t,y) = \cosh \mu y - \eta_0 \sinh \mu y$ and $\eta_0 > 1$) and the
non-$Z_2$ symmetry breaking parameter satisfies $0<f_2< (a a')^2/4\mu$
then the possible solutions for $\bar{\eta}_2$ are both negative. The
case where $f_2> (a a')^2/4\mu$ of course has no equilibrium solution. If
however, $f_2<0$ then one of the solutions will be positive, therefore
one could have two positive tension branes in equilibrium in a
semi-infinite five dimensional space-time. This
corresponds to the setup examined in~\cite{lykken} and will be discussed
in more detail later. 
Setting $\dot{\R} = \ddot{\R} = 0$ in equation (\ref{5Rdd}) leads to a
similar constraint on $p_2$,
\begin{equation}\label{pconA}
\frac{n'}{n} \;\; = \;\;
             - 2\mu \bar{\eta}_2 - 3\mu \bar{p}_2 \;+\;
                \frac{3f_2 \bar{p}_2}{\bar{\eta}_2^2 a^4},
\end{equation}
where $\bar{p}_2$ is the dimensionless pressure on the second brane at 
equilibrium. This constraint can be rewritten using equation (\ref{eta2conA}) to give,
\begin{equation}\label{pconB}
\bar{p}_2 \;\; = \;\;
      \frac{   
         \bar{\eta}_2 \left(
                            \frac{n'}{n} + 2\mu \bar{\eta}_2
                      \right) 
           }
          {
           3 \left(
                   \frac{a'}{a} - 2\mu \bar{\eta}_2
             \right)
          }\, .
\end{equation}
This demonstrates that except for a few trivial situations, the
equilibrium requirement forces $\omega_2$ to be time dependent and
therefore for the second brane to have an `unnatural' equation of
state. Note that equations (\ref{eta2conB}) and (\ref{pconB}) do
reproduce the Randall-Sundrum conditions $\eta_0 = -\eta_2 = 1$ or equivalently 
$ \kappa^2 \rho_0 = - \kappa^2 \rho_2 = \sqrt{-6 \Lambda}$, 
in the constant brane tension ($\omega_0 = \omega_2 = -1$), 
non-$Z_2$ and time independent limit as expected. 
For the specific case where the equation of state of the reference brane 
is given by $p_0 = \omega_0 \eta_0$ with $\omega_0 = -1$, one finds from 
solving the Einstein equations that 
$a'/a = n'/n$. Equation (\ref{pconB}) then becomes (using equation (\ref{eta2conA})),
\begin{equation}\label{pconC}
\bar{p}_2 \;\; = \;\;
      \frac{   
         \bar{\eta}_2 \left(
                            \frac{f_2}{\bar{\eta}_2 a^4} + 3\mu \bar{\eta}_2
                      \right) 
           }
          {
           3 \left(
                   \frac{f_2}{\bar{\eta}_2 a^4} - \mu \bar{\eta}_2
             \right)
          }\, .
\end{equation}
Therefore, at times when the non-$Z_2$ symmetric nature of the 
second brane dominates i.e. when $f_2/\eta_2 a^4 \gg \mu \eta_2$
equation (\ref{pconC}) implies that 
$\omega_2 = 1/3$. This corresponds 
to late time in a radiation dominated universe. Alternatively, if 
the second brane is approximately $Z_2$ symmetric such that 
$f_2/\eta_2 a^4 \ll \mu \eta_2$ then equation (\ref{pconC}) 
predicts that 
$\omega_2 = -1$, which corresponds to the second brane possessing a 
constant brane tension. Another interesting case where equilibrium 
positions exist is when both branes are tuned, $Z_2$-symmetric and possess 
realistic cosmologies. Here, equilibrium positions exist provided both branes 
have the same equation of state. This situation is discussed fully 
in section \ref{MRbored}. The nature of the stability of the above equilibrium 
solutions will be examined in the next section.

\subsection{Radion Fluctuations}\label{fluc}

In order to examine the nature of the equilibrium solutions found
above it is necessary to linearise the radion equations (\ref{5R1})
and (\ref{5Rdd}) around the equilibrium points. Setting 
$\dot{\R} = \delta \dot{\R}$, $\ddot{\R} = \delta \ddot{\R}$, 
$\eta_2 = \bar{\eta}_2 + \delta \eta_2$ and 
$p_2 = \bar{p}_2 + \delta p_2$ results in,
\begin{eqnarray}
\delta \left( \frac{a'}{a} \right) \; + \; \label{linR1}
   \frac{\dot{a}}{a n} \frac{\delta \dot{\R}}{n} &=& 
    \left(
          \mu  - \frac{f_2}{\bar{\eta}_2^2 a^4}
    \right) \delta \eta_2 \; - \;
    \frac{4 f_2}{\bar{\eta}_2 a^5} \delta{a}  \, , \\
\delta \left(
             \frac{n'}{n}
       \right)
  + \frac{\delta \ddot{\R}}{n^2}   
  - \frac{\dot{n}}{n^2} \frac{\delta \dot{\R}}{n} &=&
  - 2\mu \delta \eta_2 - 3\mu \delta p_2 
  + \frac{3f_2 \bar{p}_2}{\bar{\eta}_2^2 a^4}
       \left(
             \frac{\delta p_2}{\bar{p}_2} 
          -  2 \frac{\delta \eta_2}{\bar{\eta}_2}
          -  4 \frac{\delta a}{a}
       \right). \label{linRdd1}
\end{eqnarray}
In order to evaluate the first term in each of the 
equations (\ref{linR1}) and (\ref{linRdd1}) it is now assumed that
the reference brane respects $Z_2$ symmetry and possesses no Weyl tensor
component. Therefore $f_0=0$ and $\mathcal{C}=0$ or in terms of the
bulk Schwarzschild masses on either side of the reference brane:
$\C_0 = \C_1 = 0$. Since $\C_2 \not= 0$ the second brane will not in
general be $Z_2$ symmetric. This means that the metric components, which are 
obtained by solving Einstein's equations (\ref{Ein11}), now look like~\cite{bin3},
\begin{eqnarray}\label{pub101}
a(t,y) \;\; &=& \;\; a_0 ( \cosh \mu y - \eta_0 \sinh \mu y )\, , \\
n(t,y) \;\; &=& \;\; \cosh \mu y - \tilde{\eta}_0 \sinh \mu y \, ,\label{pub102}
\end{eqnarray}
where $\tilde{\eta}_0 = \eta_0 + \dot{\eta}_0/H_0$. It can now be 
explicitly shown using equations (\ref{eta2conA}), 
(\ref{pconA}), (\ref{pub101}) and (\ref{pub102}) that, 
\begin{eqnarray}
\delta \left( \frac{a'}{a} \right) \; &=& \;
\left[
      \mu^2 - \left(
                    \mu \bar{\eta}_2 \; + \;
                     \frac{f_2}{\bar{\eta}_2 a^4}
              \right)^2
\right] \delta \R  \;\; \equiv \;\; m_a^2 \delta \R \, ,\\
\delta \left( \frac{n'}{n} \right) \; &=& \;
\left[
      \mu^2 - \left(
                    - 2\mu \bar{\eta}_2 - 3\mu \bar{p}_2 \;+\;
                \frac{3f_2 \bar{p}_2}{\bar{\eta}_2^2 a^4}
              \right)^2
\right] \delta \R \;\; \equiv \;\; m_n^2 \delta \R \, ,
\end{eqnarray}
and defining the function $\bar{g}(t,\R)$ as,
\begin{equation}
\bar{g}(t,\R) \;\;\; = \;\;\; \mu \bar{\eta}_2 \; + \; 
                        \frac{f_2}{\bar{\eta}_2 a^4},
\end{equation}
equations (\ref{linR1}) and (\ref{linRdd1}) can be rewritten in the
following form:
\begin{eqnarray} \label{linR2}
&&  \HU_2 \frac{\delta \dot{\R}}{n} \;+\;
  (m_a^2 + 4\bar{g} (\bar{g} - \mu\bar{\eta}_2))\delta \R \;= \;
 (  2 \mu  - \frac{\bar{g}}{\bar{\eta}_2}  ) \delta \eta_2 \, , \\
&& \frac{\delta \ddot{\R}}{n^2}   
  - \frac{\dot{n}}{n^2} \frac{\delta \dot{\R}}{n} 
  + (m_n^2 + 12\frac{\bar{p}_2}{\bar{\eta}_2} 
          \bar{g} (\bar{g} - \mu\bar{\eta}_2))\delta \R    \;=\; \nonumber\\
&& \quad \quad \quad \quad \quad   
\quad \quad \quad \quad \quad \quad \quad
-2(\mu + 3 \frac{\bar{p}_2}{\bar{\eta}_2}
                (\bar{g} - \mu\bar{\eta}_2))
                 \delta \eta_2  
-3(2\mu - \frac{\bar{g}}{\bar{\eta}_2})\delta p_2 \, ,  \label{linRdd2}
\end{eqnarray}
where the fact that $\delta a / a = (a'/a) \delta \R$ and that
$(\dot{a}/a n) \delta \dot{\R} = \HU_2 \delta \dot{\R}$ to first order,
has been used. Thus, it can be seen that in general the radion
fluctuations are entangled with the matter fluctuations and that 
solving the linearised equations becomes non-trivial. It is possible 
however, when the fluctuations are of the adiabatic type, 
$\delta p_2 = c_2^2 \delta \eta_2$ to take the appropriate linear 
combination of equations (\ref{linR2}) and (\ref{linRdd2}) to obtain 
an equation that depends upon the radion fluctuations only. Examining 
initially the case where $\omega_0 = -1$, $p_2 = \omega_2 \eta_2$
and therefore $c_2^2 = \omega_2$, and where 
the second brane is dominated by its lack of $Z_2$ symmetry such that 
$f_2/\eta_2 a^4 \gg \mu \eta_2$, ones finds that equations 
(\ref{linR2}) and (\ref{linRdd2}) combine to give,
\begin{equation}\label{flinRdd}
\delta \R_{, \tau \tau} \;\; - \;\;
3 \omega_2 \HU_2 \delta \R_{, \tau} \;\; + \;\;
3\omega_2(1-3\omega_2) \frac{f_2^2}{\eta_2^2 a^8} \delta \R \;=\; 0,
\end{equation}
where only the dominant terms have been kept. $\tau$ is defined as 
before as the time experienced by an observer on the second brane. 
The matter fluctuations 
are now related to the radion fluctuations by the formula:
\begin{equation}
\delta \eta_2 \;\; = \;\;
-\frac{\eta_2^2 a^4}{f_2} 
     \left(
           \HU_2 \delta \R_{, \tau} \;+\;
           \frac{3f_2^2}{\eta_2^2 a^8}\delta \R
     \right).
\end{equation}
Equation (\ref{flinRdd}) can now be solved for a radiation dominated 
second brane where $\omega_2 = 1/3$. Using the approximation 
$\HU_2 \simeq -f_2/\eta_2 a^4$ and remembering that in this situation 
$f_2 < 0$, one obtains the expression,
\begin{equation}
\delta \R \;=\;
             \delta \R_i \e^{-f_2(\tau-\tau_i)/\eta_2 a^4} \;+\;
             \left(
                   4\delta \R_i + 
                   \frac{\eta_2 a^4}{\eta_{2i} f_2}
                   \delta \eta_{2i}
             \right)              
     (1-\e^{-f_2(\tau-\tau_i)/\eta_2 a^4}),
\end{equation}
where the subscript $i$ means the value of the quantity at $\tau_i$. 
Therefore it is clear that when there is a constant brane tension on 
the reference brane and a highly non-$Z_2$ symmetric second brane such 
as is the case suggested by Randall and Lyken~\cite{lykken}, an equilibrium 
solution exists only when the second brane is radiation dominated, 
however this solution for the interbrane distance or cosmological 
radion is inherently unstable.

Another case of interest is when both branes are $Z_2$ symmetric and 
as discussed in the previous section if $\omega_0 = -1$ then an 
equilibrium solution exists only when $\omega_2 =-1$ also. Adding 
three times equation (\ref{linR2}) to equation (\ref{linRdd2}) and 
neglecting all $f_2$ terms results in,
\begin{equation}\label{scalar1}
\delta \R_{, \tau \tau} \;\; + \;\;
3 \HU_2 \delta \R_{, \tau} \;\; + \;\;
4\mu^2 (1 - \eta_2^2) \delta \R \;=\; 
-\frac{\kappa^2}{6} \delta T_2,
\end{equation}
where $\delta T_2$ is the linear perturbation of the trace of the 
matter energy-momentum tensor~\cite{bin3} and is given by,
\begin{equation}
\frac{\kappa^2}{6} \delta T_2 \;\; = \;\;
    -\mu \delta \eta_2 \;+\; 3 \mu\delta p_2.
\end{equation}
It can then be seen that equation (\ref{scalar1}) is just the equation 
of motion of a scalar field coupled to the perturbed trace of the 
matter energy-momentum tensor. This therefore confirms previous results 
found in~\cite{radionwave,radionds,locallylocal}. The coefficient 
of the $\delta \R $ term in equation 
(\ref{scalar1}) demonstrates that when both branes are $dS_4$ implying 
that both $\eta_0^2 > 1$ and $\eta_2^2 > 1$, small perturbations around the 
equilibrium radion position are unstable. Alternatively, if both branes 
are $AdS_4$ and therefore $\eta_0^2 < 1$ and $\eta_2^2 < 1$ then the 
perturbations are stable. It is again possible in this case to eliminate 
the matter fluctuation terms by taking a suitable linear combination 
of equations (\ref{linR2}) and (\ref{linRdd2}). For a general $\omega_2$ 
and keeping $\omega_0 = -1$ this leads to,
\begin{equation}\label{f=0Rlin}
\delta \R_{, \tau \tau} \;\; + \;\;
(2+3\omega_2) \HU_2 \; \delta \R_{, \tau} \;\; + \;\;
3\mu^2(1+\omega_2) (1-(2+3\omega_2)\eta_2^2) \; \delta \R \;=\; 0. 
\end{equation}
The matter fluctuations are now related to the radion fluctuations by,
\begin{equation}\label{matfluc1}
\mu \delta \eta_2 \;\; = \;\;
   \HU_2 \; \delta \R_{, \tau} \; + \; 
   \mu^2(1-\eta_2^2) \; \delta \R.
\end{equation}
Assuming that both branes are $dS_4$, that $\omega_2 = -1$ and that  
$k=\C_1=\C_2=0$ ensures that $\HU_2$ is time independent and given 
by $\HU_2 = \mu^2 (\eta_2^2-1)$. It is therefore possible to solve 
equation (\ref{f=0Rlin}) for $\R(\tau)$ giving,
\begin{equation}
\delta \R \;=\;
             \delta \R_i \e^{\HU_2(\tau-\tau_i)} \;-\;
             \frac{1}{\HU_2^2}\delta \eta_{2i}    
     (1-\e^{\HU_2(\tau-\tau_i)}).
\end{equation}
where as before $\delta \R_i$ and $\delta \eta_{2i}$ are the 
initial radion fluctuations and energy density fluctuations at 
time $\tau_i$. This describes the behaviour near the equilibrium 
position and demonstrates explicitly the unstable nature of the radion 
for two $dS_4$ branes and will be compared with several numerical 
solutions discussed in the next section~\cite{bin3}.
If instead, both branes are assumed to be $AdS_4$ then it is not 
possible for $\HU_2$ to be time independent and either $k$, $\C_1$ or 
$\C_2$ must be non-zero. This leads to a recollapsing universe on the 
second brane. Again taking $\omega_2=-1$ allows 
equation (\ref{f=0Rlin}) to be written as simply,
\begin{equation}\label{Rtt}
\delta \R_{, \tau \tau} \;\; = \;\; \HU_2 \; \delta \R_{, \tau}.
\end{equation}
Examining equations (\ref{matfluc1}) and (\ref{Rtt})
it can be seen that the radion will only exhibit a pseudo-stable behaviour 
in that $\R$ will initially accelerate toward the equilibrium position 
but will only decelerate once $\HU_2$ has changed from positive to 
negative. In the section \ref{num5} the analytic results discussed here will 
be compared with complete numerical solutions of the radion 
equation for several cases of interest. Before this we go on in the next section 
to discuss the cosmological radion in terms of the static bulk perspective, in 
order to obtain an intuitive understanding of some of the numerical results that 
are presented in section \ref{num5}.

\section{The Two Brane Scenario: a Bulk Perspective}\label{bulkstuff1}

In the previous section we linearised the radion equations in order to determine 
the stability of several equilibrium positions, and therefore to obtain some idea of 
the behaviour of the radion in several cases of interest. Here we go on to briefly 
examine the 
two brane model from the perspective of a bulk observer. This will allow us to show 
that the evolution of the radion can be thought of as a competition between the 
expansion rates of the two branes, and this will help determine in which cases the branes 
collide or move apart. These predictions will be shown to be in agreement with
the numerical results presented in the next section.

We are considering purely five-dimensional spacetimes that respect the symmetries of 
three-dimensional 
homogeneity and isotropy, and that only possess a five dimensional cosmological 
constant in the bulk. It was first shown by~\cite{Kraus} that a modified 
version of Birkhoff's theorem implies that such spacetimes are in fact static. The 
bulk metric takes the general form:
\begin{equation}
\d s^2 \;=\; -  \phi(R) \: \d T^2
       \: + \: R^2 \, \gamma_{ij} \d x^i \d x^j \;+\;  \phi(R)^{-1} \d R^2,
\end{equation}
where we have defined the function $\phi(R)$ as,
\begin{equation}
\phi(R) \;\;=\;\; \left( \mu^2 R^2 \;+\; k \;-\; \frac{\mathcal{C}_i}{R^2} \right),
\end{equation}
and as before $k$ takes the values $0$, $-1$, or $1$ for flat, closed or open 
geometries, and $\gamma_{ij}$ is the corresponding metric on the unit plane, 
hyperboloid or sphere. $T$ is the time experienced by a bulk observer, and $R$ the 
coordinate of the fifth dimension, while as before the subscript $i$ is either 0, 1 or 2, 
corresponding to the bulk regions to the left, between or to the right of the branes.

The simple form of these bulk solutions is counter-balanced by the fact that in these 
coordinates both the 3-branes cannot (except in trivial cases) be stationary, and in 
general move through the bulk spacetime. In fact in the non-$Z_2$ symmetric case, the 
brane moves through two different bulk spacetimes, with for example, two different 
bulk masses either side of the brane. 
If the reference brane has a trajectory given by $R=R_0(T)$, then the 
induced metric on it is given by,
\begin{equation}\label{comeon1}
\d s^2 \;\;=\;\; -\d t^2 \;+\; R_0^2(t) \gamma_{ij} \d x^i \d x^j,
\end{equation}
where again $t$ is the time experienced by an observer on the reference brane and is related to 
the bulk time $T$ by,
\begin{equation}\label{Tt111}
\d t^2  \;\;=\;\; \left[ \phi(R_0) - \frac{1}{\phi(R_0)} \left( \frac{\d R_0}{\d T} \right)^2 
                 \right] \; \d T^2.
\end{equation}
Hence from equation (\ref{comeon1}) one can see that the apparent expansion of the brane as 
seen by such an observer is 
in fact caused by the brane's motion through the bulk~\cite{Kraus}. 
One can therefore deduce that the brane's 
position $R_0(t)$ in these bulk coordinates is proportional to the scale factor $a_0(t)$ 
of the brane 
used throughout the rest of this paper. Thus the trajectory of the brane, at least in 
terms of $t$, can be 
found from its Friedmann equation (\ref{5H_0}). In order to compare the trajectories of two such 
branes we need to solve for $R_0$ in terms of the bulk time $T$. Equation (\ref{Tt111}) 
can be rearranged to give,
\begin{equation}\label{stuff123}
\frac{\d t}{\d T} \;\;=\;\; \frac{\phi(R_0)}{(\phi(R_0) + H_0^2 R_0^2)^{1/2}},
\end{equation}
where $H_0$ is again the Hubble parameter on the reference brane given by $H_0 R_0 = d R_0 / dt$. 
This can be combined with equation (\ref{stuff123}) to show that the brane's motion in terms 
of $T$ satisfies,
\begin{equation}\label{posb1}
\frac{\d R_0}{\d T} \;\;=\;\; \frac{\d R_0}{\d t} \, \frac{\d t}{\d T} \;\;=\;\; 
\frac{ H_0 R_0 \, \phi(R_0)}{(\phi(R_0) + H_0^2 R_0^2)^{1/2}}.
\end{equation}
By replacing $R_0$ and $H_0$ with $R_2$ and $H_2$, one can obtain the equivalent equation for the 
second brane,
\begin{equation}\label{posb2}
\frac{\d R_2}{\d T} \;\;=\;\; 
\frac{ \HU_2 R_2 \, \phi(R_2)}{(\phi(R_2) + \HU_2^2 R_2^2)^{1/2}}.
\end{equation}
It must be remembered that here $R_0(T)$ and $R_2(T)$ correspond to the positions of the branes in the 
static spacetime that lies between them.
The interbrane distance in terms of these bulk coordinates is now just given by $R = R_0 - R_2$, 
where $R_0(T)$ and $R_2(T)$ can be found by solving equations (\ref{posb1}) and (\ref{posb2}). 
For example, 
if we assume the reference brane is $Z_2$ symmetric, equation (\ref{posb1}) becomes,
\begin{equation}\label{contenR12}
\frac{\d R_0}{\d T} \;\;=\;\; \frac{H_0 \phi(R_0)}{\mu |\eta_0|},
\end{equation}
which assuming that $k=\C_0 = \C_1=0$, reduces to,
\begin{equation}
\frac{\d R_0}{\d T} \;\;=\;\; \mu^2 R_0^2 \sqrt{ 1 - \frac{1}{\eta_0^2}}.
\end{equation}
If the reference brane possesses a constant dimensionless energy density $\eta_0$, then this can 
be solved to give the position of the reference brane to be,
\begin{equation}
R_0 (T) \;\;=\;\; \left[ 
                        \frac{1}{R_{0i}} \:-\: \mu^2 \sqrt{ 1 - \frac{1}{\eta_0^2}} (T-T_i)
                  \right]^{-1},
\end{equation}
where $R_{0i}$ is the position of the brane at time $T_i$. If we make similar assumptions for the 
second 
brane, in that it is also $Z_2$ symmetric ($\C_2=0$) and that $\eta_2$ is constant, then $R_2(T)$ is 
found to be of the same form,
\begin{equation}
R_2 (T) \;\;=\;\; \left[ 
                        \frac{1}{R_{2i}} \:-\: \mu^2 \sqrt{ 1 - \frac{1}{\eta_2^2}} (T-T_i)
                  \right]^{-1},
\end{equation}
where again $R_{2i} = R_2(T_i)$. Now taking the convention $R_{0i} > R_{2i}$,
which here corresponds to the reference brane being of positive tension and the second brane 
being of negative tension, we can ask under what 
conditions the branes collide. $R_0(T)=R_2(T)$ occurs at a time $T_c$ given by,
\begin{equation}
T_c \;\;=\;\; \frac{1/R_{2i} \:-\: 1/R_{0i}}{\mu (\HU_2 / |\eta_2| \:-\: H_0 / \eta_0)} \;+\; T_i
\end{equation}
however the collision must occur for positive $R_0$ and $R_2$, and $R_0(T) > 0$ only when 
$T < T_{\infty}$, 
where $T_{\infty}$ is given by,
\begin{equation}
T_{\infty} \;\;=\;\; \frac{\eta_0}{\mu H_0 R_{0i}} \;+\; T_i.
\end{equation}
Therefore a collision will occur only if $T_c < T_{\infty}$, which implies the condition,
\begin{equation}\label{cond1}
\frac{R_{2i}}{R_{0i}} \;\; > \;\; \frac{H_0 |\eta_2|}{\HU_2 \eta_0} \;\;=\;\; 
            \left( \frac{1-1/\eta_0^2}{1-1/\eta_2^2} \right)^{1/2}.
\end{equation}
So using the bulk perspective we have found the requirement for brane collision to occur in terms 
of the branes' energy densities and initial positions, in the constant brane tension case. 
This condition (\ref{cond1}), is in fact 
a specific case of the brane based equilibrium conditions (\ref{eta2conA}) and (\ref{pconA}). This 
will be discussed further in section~\ref{2AdS} where we look at the constant brane tension case from 
the brane perspective, and where the equivalence of these conditions will be demonstrated explicitly.

For more realistic cases such as where both branes possess a cosmologically evolving energy density, 
the bulk perspective is less useful. It is, in general not possible to analytically solve 
equations (\ref{posb1}) and (\ref{posb2}) in order to obtain the trajectories of the branes and 
hence it is difficult to determine whether the branes will collide. In addition one cannot easily 
determine
the conditions for equilibrium of the interbrane distance; however it must be noted that 
this equilibrium is in some sense a creation of the brane coordinates themselves.
One inevitably has to turn to 
numerical methods, and due to this in the next section we numerically integrate the radion equation, 
formulated in terms of the brane coordinates.
Note that in the brane based perspective, all the information about the branes' relative positions 
contained in equations (\ref{posb1}) and (\ref{posb2}) is 
described by one equation (\ref{5R1}), the non-linear equation for the radion. Hence we only need to 
numerically solve one non-linear equation.

The bulk perspective however, can still provide intuitive insight into the results obtained in the 
following chapter. Thinking of the evolution of the interbrane distance as a competition between 
the branes' expansion rates helps explain the rich variety of behaviour that is found.

\section{Numerical Analysis of the Non-Linear Radion}\label{num5}

Previously, the equations for the cosmological radion have been 
derived and analysed. The equilibrium points have been identified 
and the stability of these solutions examined. Unfortunately, due to 
the highly non-linear nature of the radion equation there is only so 
much one can learn from analytical methods. Hence, in this section, 
the radion equations are solved numerically in several different cases 
in order to determine the non-perturbative aspects of the cosmological 
radion. The subtle numerical problems involved are discussed, and the results are interpreted 
using the bulk perspective described in the previous section. We first review and 
then extend some of the results on $dS$ and $AdS$ branes presented in~\cite{bin3}.

\subsection{Constant Tensions on Both Branes: The $dS$ Case}\label{2AdS}

\begin{figure}
 \center
  \epsfig{file=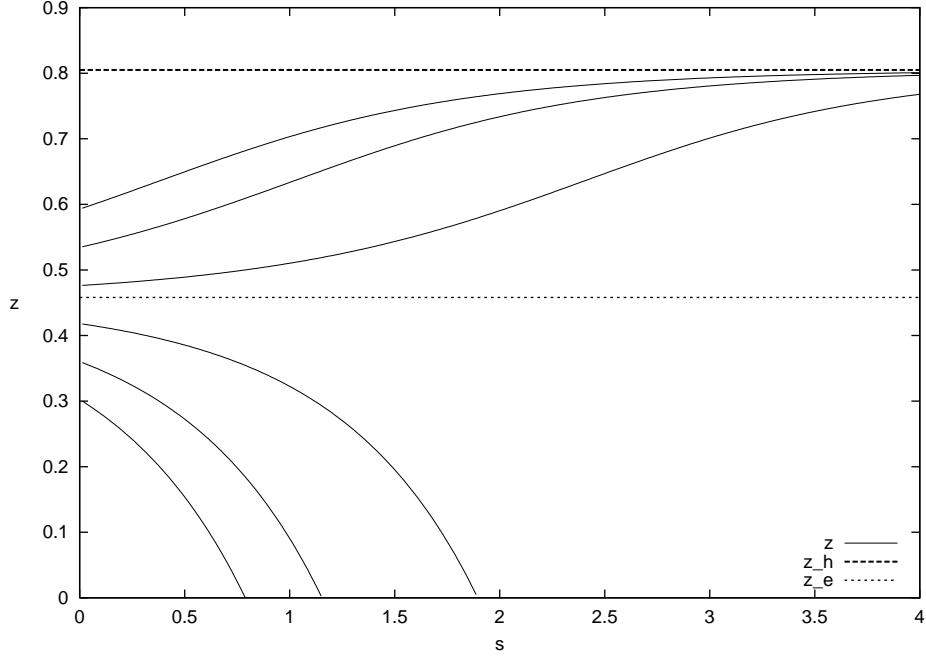,angle=270,scale=0.5}
  \caption{Both branes are $dS$ and $\eta_0 = 1.5$, $\eta_2 = -3$.}
 \label{c+1.5c-3}
\end{figure}

\begin{figure}
 \center
  \epsfig{file=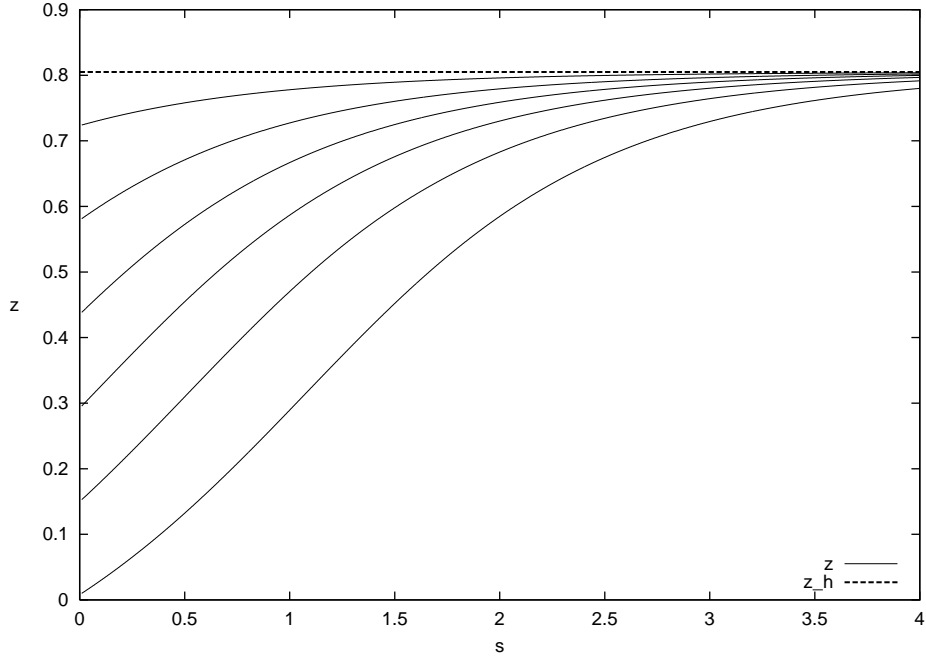,angle=270,scale=0.5}
  \caption{Both branes are $dS$ and $\eta_0 = 1.5$, $\eta_2 = -1.3$.}
 \label{c+1.5c-1.3}
\end{figure}

If the assumptions are made that the reference brane is $dS_4$ or 
equivalently that $\eta_0>1$, and that $\C_0=\C_1=0$ implying that 
the brane possesses $Z_2$ symmetry and has no Weyl tensor component, 
then the brane based metric components can be written in the simple form,
\begin{eqnarray}\label{a5}
a(t,y) &=& a_0(t) (\co - \eta_0 \si) \;=\;
             a_0(t) \sqrt{\eta_0^2 -1} \sinh{\mu(y_h - |y|)}\, , \\
n(y)   &=&  (\co - \eta_0 \si) \;=\;
             \sqrt{\eta_0^2 -1} \sinh{\mu(y_h - |y|)}\, ,\label{n5}
\end{eqnarray}
where $y_h$ denotes the position of the coordinate singularity defined 
by $a(t,y)=0$ and is given by,
\begin{equation}\label{horizon1}
\mu y_h \;\;=\;\; \tanh^{-1}\frac{1}{\eta_0}.
\end{equation}
Replacing the metric components (\ref{a5}) and (\ref{n5}) into 
equations (\ref{eta2conA}) and (\ref{pconA}) to determine the 
equilibrium radion position denoted by $y_e$, one finds as stated 
before that the equation of state on the second brane must be of the 
form $p_2 = - \eta_2$ and that,
\begin{equation}\label{equil1}
\mu y_e \;\;=\;\; \tanh^{-1}\frac{1}{\eta_0} \;+\;
\tanh^{-1}\frac{1}{\eta_2}.
\end{equation}
In order to numerically solve the radion equation here and in the 
following sections, it is useful to convert to the following 
dimensionless variables,
\begin{equation}
z \;=\; \mu \R, \quad z_{h,e} \;=\; \mu y_{h,e} \quad
s \;=\; \mu t, \quad
h_0 \;=\; \frac{H_0}{\mu}, \quad h_2 \;=\; \frac{\HU_2}{\mu}.
\end{equation}
Replacing the expressions for the metric coefficients (\ref{a5}) and 
(\ref{n5}) into the radion equation (\ref{5R1}) and changing to the 
above dimensionless variables gives,
\begin{equation}\label{z5}
z_{,s} \;\;=\;\; h_0 \sinh(z_h - z)
            \frac{
                  \cosh(z_h - z) + h_2 \eta_2 \sinh^2(z_h - z)
                 }
                 {
                  1 + \eta_2^2 \sinh^2(z_h - z)
                 },
\end{equation}
where it should be remembered that here $h_0^2 = \eta_0^2 -1$, and 
setting $k = 0$ gives $h_2^2 = \eta_2^2 -1$. Since the energy densities 
of both branes are by assumption constant in time, the evolution of 
the radion is completely embodied by equation (\ref{z5}) and all that 
is required to solve it is the initial radion position denoted by $z_i$. 
Numerical solutions were generated for two different 
situations, each with varying initial radion positions and the results 
are shown in figures \ref{c+1.5c-3} and \ref{c+1.5c-1.3}. 
Figure \ref{c+1.5c-3} 
corresponds to the case where $\eta_0 = 1.5$ and $\eta_2 = -3$ and shows 
the trajectories of six different initial radion positions distributed 
around the equilibrium position $z_e$. Equations (\ref{horizon1}) 
and (\ref{equil1}) show that for this choice of brane tensions the 
equilibrium and singularity positions are $z_e\simeq 0.458 $ and 
$z_h\simeq 0.805$ which correspond to the first and second dotted lines 
of figure \ref{c+1.5c-3}. As expected from the above analysis, 
the equilibrium 
position is found to be unstable: initial positions satisfying 
$z_i < z_e$ lead to the second brane colliding with the 
reference brane; as opposed to initial positions satisfying $z_i>z_e$ 
when the second brane 
asymptotically `freezes out' at $z_h$ as seen by an 
observer on the reference brane. Consideration of equations (\ref{z5}) 
and (\ref{tau5}) demonstrates that crossing $z_h$ only takes a 
finite amount of proper time on the second brane.


The situation where $\eta_0 = 1.5$ and $\eta_2 = -1.3$ was investigated 
and the results are shown in figure \ref{c+1.5c-1.3}. Referring to the 
expression for the equilibrium position given by equation (\ref{equil1}), 
it can be seen that if both branes are $dS$ with for example $\eta_0$ 
positive and $\eta_2$ negative, then $|\eta_0| > |\eta_2|$ implies 
that $z_e <0$. Due to the $Z_2$ symmetry this means that there is no 
physical equilibrium position for the radion in this setup. This is 
confirmed by figure \ref{c+1.5c-1.3} where every choice of initial 
position for the radion leads to the second brane freezing out at 
$z_h$, where $z_h \simeq 0.805$ as in the previous example.
This behaviour can be better understood from a bulk 
observer's perspective. In the bulk coordinate system, both branes are 
moving through a static background. Their 
trajectories from this point of view are solely dependent upon their 
expansion rates or Hubble constants; increasing the magnitude 
of $\eta_2$ increases $\HU_2$ and hence improves the chances of the 
second brane `catching' the reference brane. Alternatively, if as in 
figure \ref{c+1.5c-1.3}, the magnitude of $\eta_0$ is greater than that 
of $\eta_2$, it will be impossible for the branes to collide. 
In section~\ref{bulkstuff1} we used the bulk perspective to solve for the 
trajectories of two constant tension $dS$ branes, and derived the condition for collision
(\ref{cond1}): $R_{2i}/R_{0i} > H_0 |\eta_2| / \HU_2 \eta_0 $. This can be rewritten 
using the fact that $R_0$ and $R_2$ are proportional to $a_0(t)$ and $a(t,z)$ 
respectively, which implies,
\begin{equation}
\frac{\alpha a(t,z)}{ a_0(t)} \;\; > \;\; \frac{H_0 |\eta_2|}{\HU_2 \eta_0},
\end{equation}
where we have scaled the three spatial dimensions $x^i$ so that $R_0 = a_0$ and 
therefore $R_2 = \alpha a(t,z)$ where $\alpha$ is some positive constant discussed below.
Again using the explicit expression for the metric element $a(t,z)$ given by equation 
(\ref{a5}), this becomes,
\begin{equation}
\sinh (z_h - z) \;\;>\;\; 
         \frac{|\eta_2|}{\alpha \eta_0} \: \frac{1}{(\eta_2^2 -1)^{1/2}}.
\end{equation}
Now assuming that $\alpha = |\eta_2| / \eta_0$, one finds that the branes will collide if the 
initial $z$ satisfies,
\begin{equation}
\tanh (z_h - z) \;\;>\;\; \frac{1}{|\eta_2|}, 
\end{equation}
from which one can recover the equilibrium condition equation (\ref{equil1}), derived from 
the brane perspective. One can check that $\alpha$ does indeed take the required form, as 
it must in order 
for the two perspectives to agree, by checking the explicit transformation between 
bulk and brane coordinates as detailed in~\cite{muko}.

The bulk 
perspective is very useful for understanding the behaviour of the 
non-perturbative cosmological radion and will be discussed further. It 
must be stressed that a time independent bulk solution will not be 
possible for scenarios that contain more general matter in the bulk, 
such as for example scalar fields or dilaton fields. This is another 
reason for using brane based coordinates to examine the radion.

\subsection{Constant Tensions on Both Branes: The $AdS$ Case}

\begin{figure}
 \center
  \epsfig{file=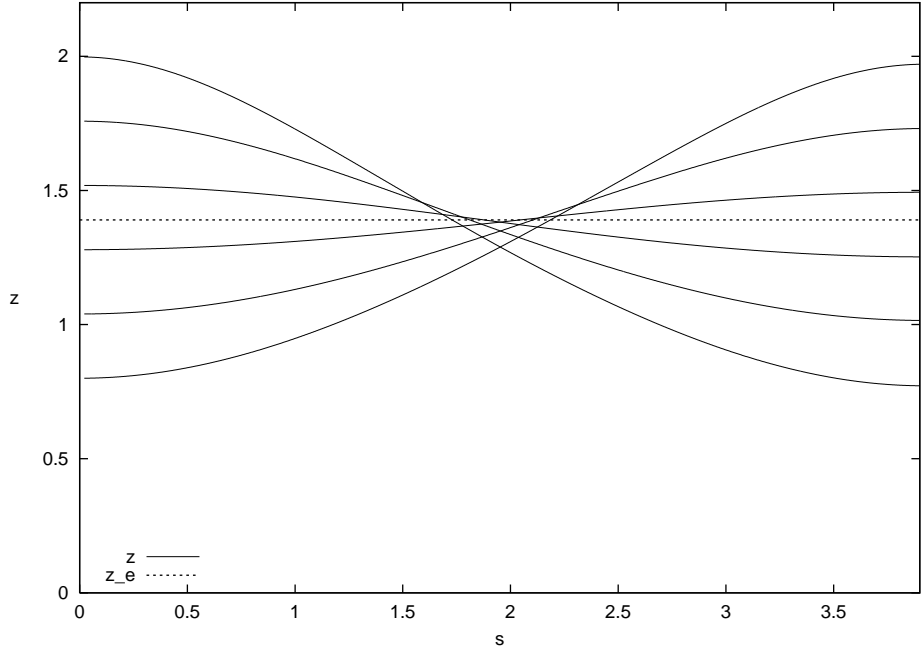,angle=270,scale=0.5}
  \caption{Both branes are $AdS$ and  $\eta_0 = 0.6$, $\eta_2 = 0.6$.}
 \label{c+0.6c+0.6}
\end{figure}

\begin{figure}
 \center
  \epsfig{file=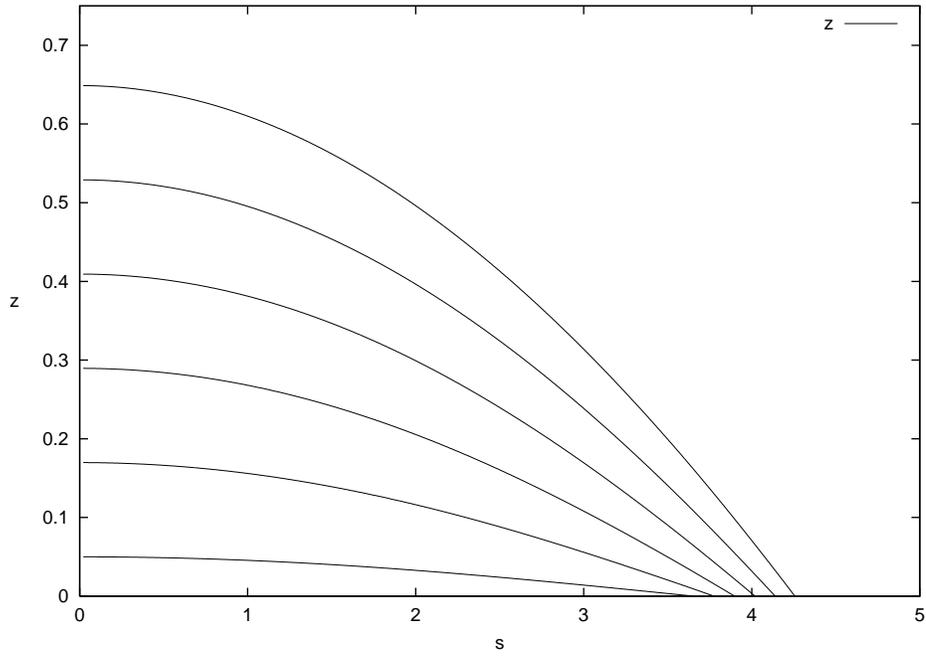,angle=270,scale=0.5}
  \caption{Both branes are $AdS$ and $\eta_0 = 0.9$, $\eta_2 = -0.9$.}
 \label{c+0.9c-0.9}
\end{figure}

In this section the radion dynamics are investigated for the case of 
two $AdS_4$ branes, that is two branes such that $|\eta_0|<1$ and 
$|\eta_2|<1$. Again it is assumed that both branes are
$Z_2$-symmetric and that they possess no Weyl tensor component such
that $\C_0=\C_1=\C_2=0$. This implies that one must take $k=-1$ to get
a consistent solution to the Einstein equations. The metric
coefficients take the form,
\begin{eqnarray}\label{a6}
a(t,y) &=& a_0(t) (\co - \eta_0 \si) \;=\;
             a_0(t) \sqrt{1 - \eta_0^2} \cosh{(|z|- z_m)}\, , \\
n(y)   &=&  (\co - \eta_0 \si) \;=\;
             \sqrt{1 - \eta_0^2} \cosh{\mu(|z| - z_m)}\, ,\label{n6}
\end{eqnarray}
where the dimensionless coordinate $z=\mu y$ has been used as before,
and $z_m$ which corresponds to the minimum of $a(s,z)$ and $n(s,z)$ is
given by,
\begin{equation}
z_m \;\;=\;\;
         \tanh^{-1} \eta_0.
\end{equation}
From equation (\ref{5R2}) it is seen that since $a(s,z)$ possesses a
minimum, it is possible to have two positive tension branes 
in a compactified five-dimensional spacetime~\cite{s102}. The radion 
equation (\ref{5R1}) now looks like,
\begin{eqnarray}\label{AdSrad}
\!\!\!\!\!\!\! z_{,s} &=& K_0^2 \cosh(z - z_m) \;\; \times  \\
    &&        \frac{
                  h_0 \sinh(z - z_m) \pm  
                  \eta_2 \cosh(z - z_m)
                  \sqrt{ h_0^2 + K_0^2
                         (1-K_2^2\cosh^2(z-z_m))}
                 }
                 {
                  h_0^2 \; + \; \eta_2^2 K_0^2
                  \cosh^2 (z-z_m)
                 },
\end{eqnarray}
where $K_0= \sqrt{1-\eta_0^2}$ and $K_2 = \sqrt{1-\eta_2^2}$. Using 
equation (\ref{5H_0}) the dimensionless Hubble parameter on the reference 
brane is now given by,
\begin{equation}
h_0 \;\; = \;\; -K_0^2 \; + \; \frac{1}{a_0^2} \, ,
\end{equation}
which can be solved to give $a_0(t) = \sin (K_0 s) / \mu K_0$ and therefore 
$h_0 = K_0 /\tan(K_0 s)$ which is all that is needed in order to 
numerically solve equation (\ref{AdSrad}). It must be noted that, when 
generating a solution to equation (\ref{AdSrad}) one must be careful to 
ensure that the sign of the square root is changed whenever the argument of 
the square root vanishes i.e. when $h_2=0$. This has to be the case and can 
be seen on examination of equation (\ref{5R1}). The equilibrium position 
can now be shown to be,
\begin{equation}
z_e \;\; = \;\; \tanh^{-1} \eta_0 \; + \; \tanh^{-1} \eta_2.
\end{equation}
Figures \ref{c+0.6c+0.6} and \ref{c+0.9c-0.9} show the radion trajectories 
that were numerically generated for two distinct situations. The first case 
that was examined, corresponding to figure \ref{c+0.6c+0.6}, was where 
$\eta_0 = 0.6$ and $\eta_2 = 0.6$ which implies choosing the positive root 
in equation (\ref{AdSrad}). The initial 
radion positions are distributed about the equilibrium point 
$z_e = 1.39$ which is represented by the dotted line. As expected from the 
analysis in section~\ref{fluc}, it is seen that the the radion displays 
a pseudo-stable behaviour - initially the interbrane distance accelerates 
toward $z_e$, but it only begins to decelerate once $h_2=0$ which does 
not necessarily correspond to $z$ crossing the equilibrium point. It is this 
fact combined with the warped nature of $n(s,z)$ that causes different 
trajectories to reach $h_2=0$ at different points and hence leads to the 
asymmetrical (in the $z$ direction) appearance of figure \ref{c+0.6c+0.6}.
Figure \ref{c+0.9c-0.9} shows several radion trajectories when the second brane 
possesses a negative tension. Specifically, the brane tensions have been 
chosen to be $\eta_0 = 0.9$ and $\eta_2 = -0.9$ which implies that $z_e =0$. 
As expected, this leads to the second brane colliding with the first for 
any given initial radion position.

\subsection{Radiation and Matter on the Reference Brane}

\begin{figure}
 \center
  \epsfig{file=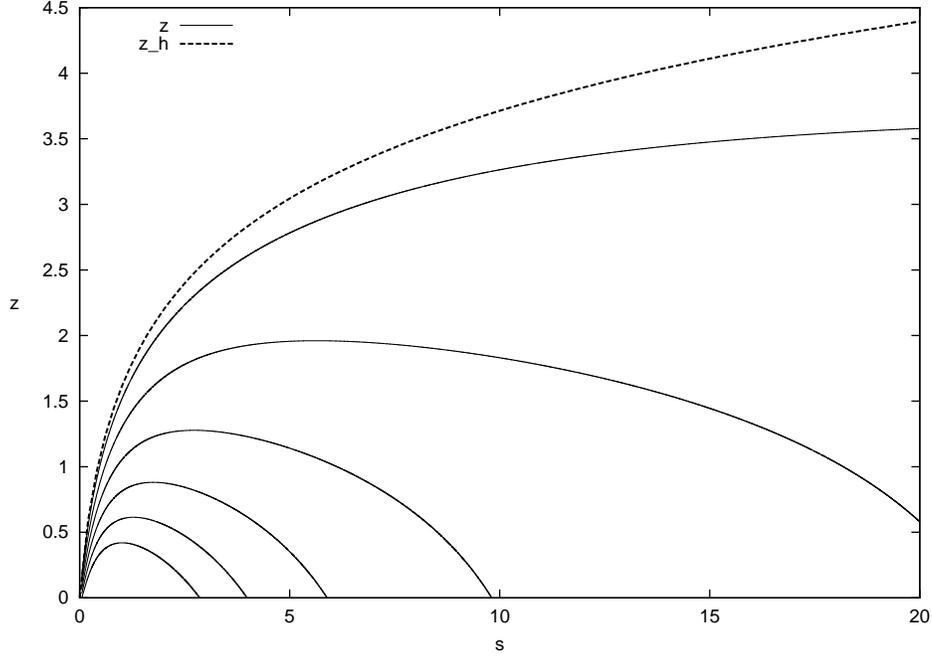,angle=270,scale=0.5}
  \caption{Radiation on the reference brane, $\eta_2 = -1.2$ on
   the second brane.}
 \label{r+c-1.3}
\end{figure}

We now turn to a more interesting situation where there exists a realistic 
cosmology on the reference brane.  We assume that the 
dimensionless energy density $\eta_0$ can be decomposed into two parts~\cite{csaki}: 
a constant brane tension and a `physical' energy density denoted by 
$\eta_{0p}$ such that,
\begin{equation}\label{one}
\eta_0 \;\; = \;\; 1 \; + \; \eta_{0p},
\end{equation}
and similarly the dimensionless pressure is decomposed into,
\begin{equation}
p_0 \;\; = \;\; - 1 \; + \; p_{0p}.
\end{equation}
This is done as before to ensure that the reference brane undergoes 
standard cosmology at late times or more specifically that the Friedmann 
equation has the form,
\begin{equation}\label{Fri5}
H_0^2 \;\; = \;\; \mu^2 \eta_{0p}^2 + 2 \mu^2 \eta_{0p} \, ,
\end{equation}
so that $H_0^2 \propto \eta_{0p}$ at late times, where we have assumed that both 
branes are $Z_2$ symmetric, possess no Weyl tensor component and that $k=0$. 
If the reference brane now has an equation of state $p_{0p} = \omega_0 \eta_{0p}$, 
equations (\ref{pub101}) and (\ref{pub102}) show that 
the metric coefficients in terms of $s$ and $z$ will now be given by,
\begin{eqnarray}\label{arad}
a(s,z) &=& a_0 (\cosh z - (1+\eta_{0p}) \sinh z) \, ,\\
n(s,z)   &=&  \cosh z - (1-(2+3\omega_0)) \eta_{0p} \sinh z \, ,\label{two}
\end{eqnarray}
where now both $a_0$ and $\eta_{0p}$ are functions of the dimensionless time 
parameter $s$. The Friedmann equation (\ref{Fri5}) can now be solved for 
$\eta_{0p}$ using the brane energy conservation equation,
\begin{equation}
\dot{\eta}_0 \;\;=\;\; -3 H_0 (\eta_0 \:+\: p_0),
\end{equation}
which leads to,
\begin{equation}\label{etap}
\eta_{0p}(s) = \frac{1}{\frac{1}{2}q_0^2 s^2 + q_0 s},
\end{equation}
where we have defined $q_0 = 3(1+\omega_0)$. Using this expression 
for $\eta_{0p}(s)$, the position of the 
coordinate singularity defined by $a(s,z_h)=0$ is now given by the 
simple formula,
\begin{equation}\label{hor5}
z_h \;\; = \;\; \tanh^{-1} \left(\frac{1}{1+\eta_{0p}} \right)\;\; = \;\;
                \ln( q_0 s + 1).
\end{equation}
We now assume that the reference brane is in a radiation dominated phase 
in that $\omega_0 = 1/3$, $q_0=4$ and therefore $\eta_{0p} \propto a_0^{-4}$
and that the second brane has a constant brane tension. 
It is now possible, using equations (\ref{one}), (\ref{Fri5}), (\ref{arad}),
(\ref{two}) and (\ref{etap}) to numerically solve the radion 
equation~(\ref{5R1}). This was done using several different 
initial conditions and the results are shown in figure 
\ref{r+c-1.3}, along with the time dependent position of $z_h$ 
denoted by the dotted line and given by equation (\ref{hor5}). 
The six initial radion positions were 
distributed evenly between $z=0$ and $z=z_h$. In figure \ref{r+c-1.3} the 
radion positions are graphed against time for a second brane tension 
$\eta_2 = -1.2$, and as can be seen, all the radion trajectories lead 
to the two branes colliding. Trajectories that initially begin closer to 
$z_h$ take longer to return to the reference brane as would 
be expected.

\begin{figure}
\center
\epsfig{file=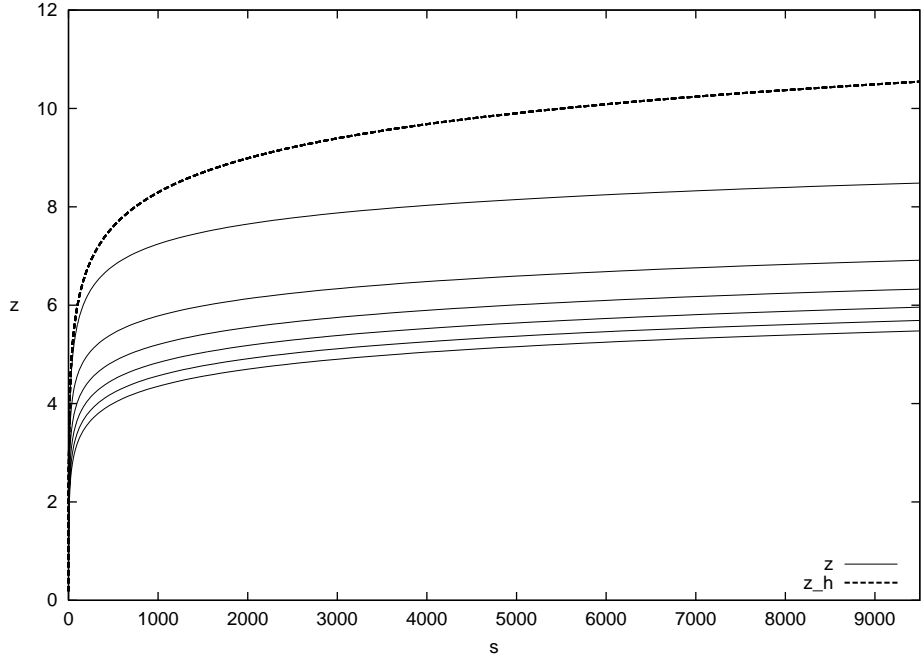,angle=270,scale=0.5}
\caption{Radiation on the reference brane with a tuned ($\eta_2 = -1$) second brane.}
\label{Seven}
\end{figure}

Similar results were also obtained for various values of the second brane tension. 
It was found that the radion trajectories exhibit the same 
general behaviour, but the second 
brane would return and collide with the reference brane sooner/later if the magnitude of 
$\eta_2$ was greater/lesser.
An interesting case is shown in figure~\ref{Seven}, where the reference brane is again radiation 
dominated, however the second brane is now a tuned or critical brane with $\eta_0 = -1$. 
The initial radion positions were spread between $z=0$ and $z=0.95z_h$ at $s=0.05$
and the coordinate singularity 
moves away from the reference brane as before with $z_h = \ln (4s+1)$. In this 
situation the second brane does not return and collide with the reference brane, 
instead the interbrane distance increases with logarithmic behaviour similar to that 
of the coordinate singularity at $z_h$.
If instead, it is assumed that the reference brane is in a matter 
dominated phase so that $\omega_0 = 0$, $q_0=3$ and therefore 
$\eta_{0p} \propto a_0^{-3}$, the dynamics of the interbrane distance will again be 
altered. The coordinate singularity is now at $z_h = \ln (3s+1)$, and it was found that 
if the reference brane is matter dominated, the second brane takes longer to return.  

This general radion behaviour when the reference brane has a more realistic 
cosmology can be understood in terms of the evolution of the brane energy 
densities and by viewing the situation from a bulk observers perspective.
Initially $\eta_{0p}$ is very large causing $a_0$ to increase rapidly
($a_0 \propto t^{1/4}$) and therefore the reference brane will be moving 
rapidly through the static bulk. The second brane on the other hand, only 
has a low constant energy density $\eta_2$, and will begin by moving slowly, 
hence the interbrane distance will initially increase. At late times, 
$\eta_{0p}$ tends to zero and the reference brane is much 
slower compared to the second brane which now has 
$a_2 \propto \exp \HU_2 \tau$ and therefore the interbrane distance will 
decrease and eventually the branes will collide. This means that the time 
taken to collide is decreased by having a larger $\eta_2$, in agreement with 
the above results. Alternatively, a matter dominated reference brane will 
maintain a higher `speed' at late times and therefore will increase the time 
taken to collide. In figure~\ref{Seven}, where the tuned second brane never returns, 
we have that $\HU_2 =0$, and therefore from the bulk 
observers point of view the second brane is stationary. Since our brane is 
expanding (albeit with a continually decreasing rate), it moves away from the second brane and
hence the interbrane distance increases monotonically. All these features of the radion's 
behaviour will be discussed further in the next section when we go on to investigate the 
time taken for the branes to collide.

\subsubsection{The Time Taken for the Branes to Collide}

\begin{figure}
\center
\epsfig{file=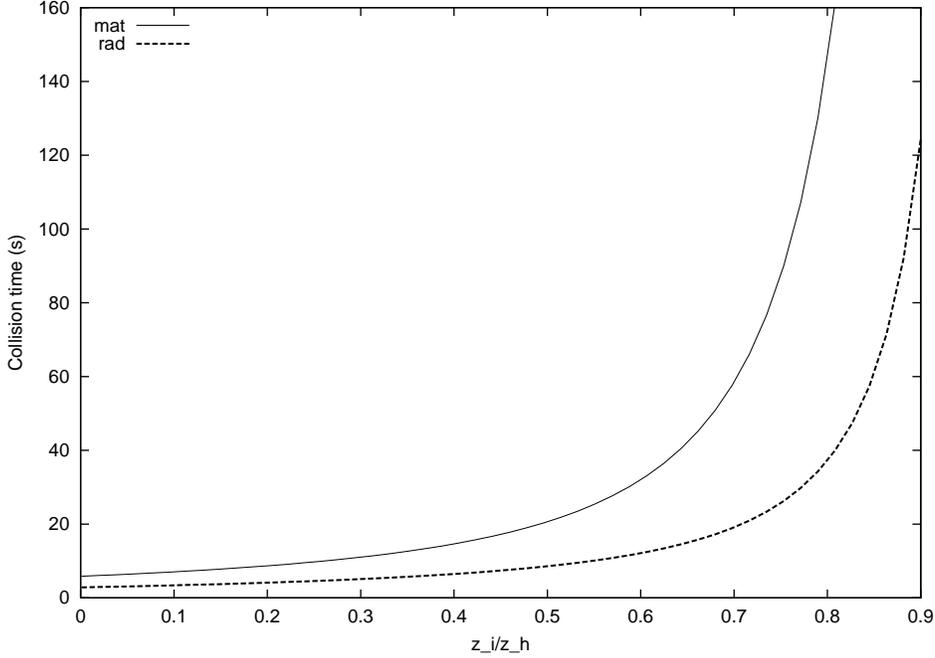,angle=270,scale=0.5}
\caption{Time taken for branes to collide for varying initial $z$}
\label{One}
\end{figure}

In order to obtain an intuitive understanding of the nature of the radion it is useful 
to investigate explicitly $s_{col}$, the dimensionless time taken for the branes to collide. $s_{col}$ 
depends on many variables including the initial conditions, the equations of state on 
either brane, the curvature and non-$Z_2$ symmetry of the branes and the proximity of the 
second brane to $z_h$. Here we have restricted our analysis to the situations presented 
in the previous section, where the reference brane possesses a realistic cosmology and the second 
brane has a constant negative brane tension. The dependence of the collision time on 
the initial radion position $z_i$, when the reference brane is either radiation or matter 
dominated was numerically determined and the results are shown in figure~\ref{One}. Firstly 
one can see that for all values of $z_i/z_h$, the branes collide much sooner for a 
radiation as opposed to a matter dominated reference brane. As discussed in the previous section,
this is due to the matter dominated brane moving through the bulk with a greater `speed' than that 
of the radiation dominated brane. 
Secondly, as the initial radion position increases, the collision time also increases,
becoming singular as $z_i \rightarrow z_h$. The reason for this is that $a(s,z)\rightarrow 0$ 
at $z_h$, and therefore the initial scale factor of a brane close to $z_h$ tends to zero also. Since 
the second brane has to expand from $a(s,z_i)$ to $a_0(s_{col})$ in order for collision to occur, 
a vanishing $a(s,z_i)$ implies $s_{col} \rightarrow \infty$.

\begin{figure}
\center
\epsfig{file=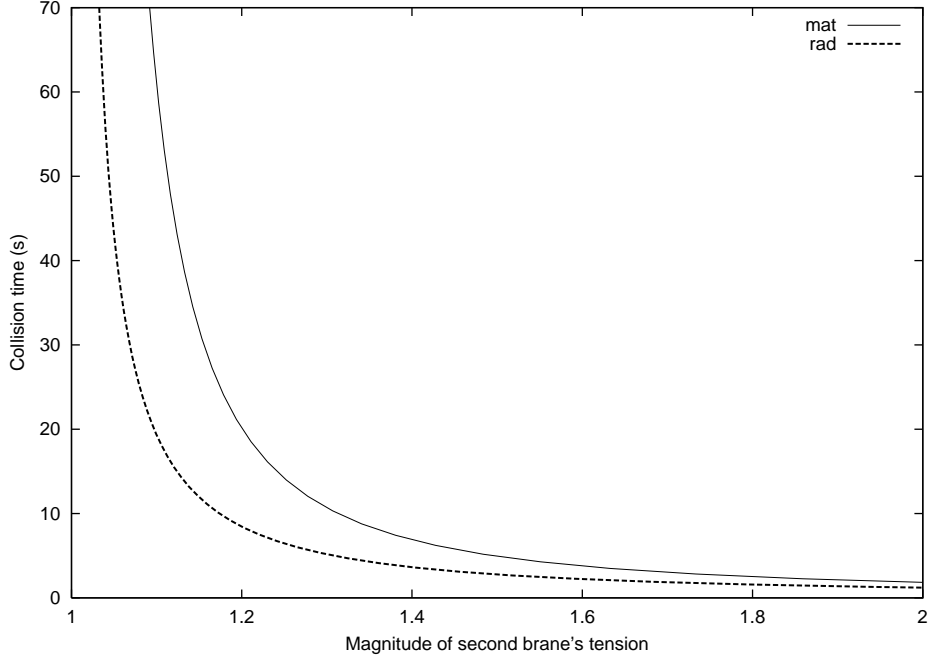,angle=270,scale=0.5}
\caption{Time taken for branes to collide for varying $\eta_2$}
\label{Two}
\end{figure}

The effect on the collision time of varying the tension of the second brane was investigated, 
and the results are shown in figure~\ref{Two}. A similar effect to figure~\ref{One} is seen in 
that radiation domination implies faster collision. In addition, it can be seen that increasing the 
magnitude of $\eta_2$ leads to the second brane moving faster, which in turn decreases $s_{col}$. 
As can be seen from figure~\ref{Seven}, when $\eta_2 \rightarrow -1$ then 
$s_{col} \rightarrow \infty$, since the second brane is effectively stationary from the bulk 
point of view.

\subsubsection{The Ekpyrotic and Cyclic Models}

The above discussion of the collision times of the two branes highlights an important 
feature of the well known ekpyrotic and cyclic models. In the Ekpyrotic 
universe~\cite{s43,s44,s45,s48,tol1,tol2} it was 
proposed that the supposed beginning of our four-dimensional universe is in fact 
the result of a collision between a boundary and a bulk brane. 
Some of the energy of the collision would be converted into a hot big bang, and it was shown 
how one therefore no longer needs to incorporate inflation to explain the current observational 
data. The cyclic model~\cite{s46,s47} went one step further: it was suggested that the big bang 
was due to the 
collision of two boundary branes, and that these branes would collide periodically every several
trillion years. This would also explain the apparent accelerated expansion of the universe 
we observe today. 

Both these models, however, needed to employ a potential between the branes in order for the 
brane collisions to occur at the correct times. Although these potentials were no more complex 
than those employed in the most recent inflationary theories, they were to say the least ad hoc.
We have investigated throughout this paper the {\it{natural}} behaviour of cosmologically 
realistic branes moving in a five-dimensional $AdS$ bulk, and as can be seen from the above 
results, we find that brane collisions (at least in this model) are commonplace. We also 
however, find that the collision time will not be significantly greater than the inverse 
five-dimensional Plank mass\footnote{A fact which can be inferred by 
dimensional analysis.} ($\simeq 30 TeV$ or above), and therefore to delay 
brane collisions to the extent proposed in the Ekpyrotic and Cyclic models, a certain 
amount of fine tuning must be involved. Taking this point further, in general one might 
expect a five-dimensional model of this kind to contain a large number of branes, each 
containing different equations of state. These would necessarily be colliding over 
timescales of around $ M_5^{-1}$, so in the case of a Cyclic universe the main question to answer 
would be why our brane has not hit anything else in the last thirteen billion years.

\subsection{Radiation and Matter on Either Brane}\label{MRbored}

We now turn to the interesting case where both branes are $Z_2$ symmetric and 
have time dependent energy densities $\eta_0(t)$ and $\eta_2(t)$. This allows us 
to investigate the behaviour of the radion when both branes are either radiation 
or matter dominated. There are, however, some subtleties as defining a time dependent 
energy density on the second brane leads in most cases to 
$\eta(t \rightarrow 0) = -\infty$, which could lead to some phenomenological 
problems. 

Assuming that both brane energy densities can be decomposed into a constant 
brane tension component and a time dependent `physical' component we have,
\begin{equation}
\eta_0 \;\; = \;\; 1 \; + \; \eta_{0p},\quad p_0 \;\; = \;\; - 1 \; + \; p_{0p},
\end{equation}
for the reference brane, and,
\begin{equation}\label{assum1}
\eta_2 \;\; = \;\; -1 \; - \; \eta_{2p},\quad p_2 \;\; = \;\;  1 \; - \; p_{2p},
\end{equation}
for the second brane where $\eta_{2p}$ and $p_{2p}$ are positive quantities. 
We have assumed that $\C_0 = \C_1 = \C_2 = k = 0$, and we
want to obtain the Randall-Sundrum two brane model at late times, therefore the
assumed form of $\eta_2$ and $p_2$ given by equation (\ref{assum1}) is the only 
possible choice. 

The junction conditions across the second brane give the energy density conservation 
equation:
\begin{equation}
\dot{\eta}_2 \;\;\;=\;\;\; -3 H_2\:(\eta_2 \;+\; p_2),
\end{equation}
which in terms of $\eta_{2p}$ takes the same form,
\begin{equation}\label{RMcon}
\dot{\eta}_{2p} \;\;\;=\;\;\; -3 H_2\:(\eta_{2p} \;+\; p_{2p}).
\end{equation}
If the second brane has an equation of state of the form $p_{2p} = \omega_2 \eta_{2p}$ 
then we can use equation (\ref{RMcon}) to write,
\begin{equation}\label{2ppp}
\eta_{2p} \;\; = \;\; \left(
                            \frac{\gamma_2}{a_2}
                     \right)^{3(1+\omega_2)}
          \;\; = \;\; \left(
                            \frac{\gamma_2}{a_2}
                     \right)^{q_2},
\end{equation}
where we have defined $q_2 = 3(1+\omega_2)$. A similar argument for the reference 
brane gives also,
\begin{equation}\label{0ppp}
\eta_{0p} \;\; = \;\; \left(
                            \frac{\gamma_0}{a_0}
                     \right)^{3(1+\omega_2)}
          \;\; = \;\; \left(
                            \frac{\gamma_0}{a_0}
                     \right)^{q_0},
\end{equation}
which was implicitly used in the previous two sections. $\gamma_0$ and $\gamma_2$ are 
simply constants relating the energy densities and scale factors on either brane.

One can now ask whether an equilibrium solution for the interbrane distance exists 
in this situation. Using the conditions for equilibrium given by 
equations (\ref{eta2conA}) and (\ref{pconA}) we can write the equation for 
the possible equilibrium point $z_e$ in the same form as equation (\ref{equil1}),
\begin{equation}
z_e \;\;\; = \;\;\; \tanh^{-1} \frac{1}{(1+\eta_{0p})} \;\; + \;\;
                    \tanh^{-1} \frac{1}{(-1-\eta_{2p})}.
\end{equation}
Replacing $\eta_{0p}$ and $\eta_{2p}$ using equations (\ref{2ppp}) and (\ref{0ppp}),
and rearranging gives,
\begin{equation}\label{123}
z_e \;\; = \;\; \frac{1}{2} (q_0 - q_2) \ln{a_0} \; - \;
                \frac{1}{2} q_2 \ln{\left(\frac{a_2}{a_0} \right)} \; + \;
                \frac{1}{2} \ln{\left(\frac{\gamma_2^{q_2}}{\gamma_0^{q_0}}\right)}.
\end{equation}
This shows that in general the required expression for $z_e$ is time dependent 
(since both $a_0$ and $a_2$ depend on $t$) 
and that therefore no equilibrium position exists. If, however, we examine the late 
time behaviour of this equation then we can use the fact that $a_2/a_0 = 
\cosh{z_e} - \eta_0 \sinh{z_e} \approx \exp{(-z_e)}$ to first 
order\footnote{Technically we have had to assume both that $t\gg 1$ and that $z_e$ 
is not significantly close to the coordinate singularity at 
$z_h = \tanh^{-1}(1/\eta_0)$.}. Equation (\ref{123}) can now be arranged to give,
\begin{equation}\label{zeee}
z_e \;\; = \;\; \frac{1}{q_2 - 2} 
        \left[ 
              (q_2 - q_0) \ln{a_0} \;+\;
        \ln{\left( \frac{\gamma_0^{q_0}}{\gamma_2^{q_2}}\right)}
        \right]
\end{equation}
This demonstrates that there will be an equilibrium solution at late times if 
and only if $q_0 = q_2$ i.e. if both branes have the same equation of state. 
Therefore if both branes are radiation dominated, $q_0 = q_2 = 4$ and 
$z_e = 2\ln{(\gamma_0/\gamma_2)}$, however, if they are matter dominated then 
$q_0 = q_2 = 3$ and $z_e = 3\ln{(\gamma_0/\gamma_2)}$. 

We now investigate the stability of such solutions by examining 
equation (\ref{linR2}), which in dimensionless variables and with $f_2=0$ looks 
like,
\begin{equation}\label{zstab1}
h_2 \; \delta z_{,s_2} \;\; + \;\; (1 - \bar{\eta}_2^2) \; \delta z \;\; = \;\;
                     \delta \eta_2.
\end{equation}
where we have defined $s_2 = \mu \tau$, the dimensionless time experienced by 
an observer on the second brane. Using the fact that we have set 
$\eta_2 = -1 - \eta_{2p} = -1 - (\gamma_2 / a)^{q_2}$ gives,
\begin{equation}\label{zstab2}
\delta \eta_2 \;\; = \;\; q_2 \: \left(
                      \frac{\gamma_2}{a}
                             \right)^{q_2} \: \frac{a'}{a} \: \delta z
\;\; = \;\; q_2 \: \bar{\eta}_{2p} \: (-1 - \bar{\eta}_{2p}) \: \delta z,
\end{equation}
where we have used the fact that at equilibrium $a'/a = \bar{\eta}_2$. Combining 
equations (\ref{zstab1}) and (\ref{zstab2}) and expressing everything in terms of 
$\bar{\eta}_{2p}$ leads to,
\begin{equation}\label{flucp2}
\delta z_{,s_2} \;\; = \;\;
              - \left(
                      \frac{ \bar{\eta}_{2p}}{\bar{\eta}_{2p} + 2}
               \right)^{1/2}
                \left( 
                     q_2 - 2 \,+\, (q_2 - 1)\bar{\eta}_{2p}
               \right) \: \delta z.
\end{equation}
This shows that the fluctuations around $z_e$ will be stable as long as $q_2>2$. The 
explicit expression for $\eta_{2p}$ in terms of the time experienced by an observer on
the second brane $s_2$ can be found from the Friedmann equation (\ref{5H_3}) combined 
with equation (\ref{2ppp}) and takes the same form as equation (\ref{etap}),
\begin{equation}\label{etap22}
\eta_{2p}(s_2) = \frac{1}{\frac{1}{2}q_2^2s_2^2 + q_2 s_2}.
\end{equation}
Inserting this into the fluctuation equation allows us to solve for $\delta z(s_2)$ 
giving,
\begin{equation}
\delta z(s_2) \;\; = \;\;  \frac{A(q_2 s_2 +1)}
                              {\left[
                                     q_2 s_2 (q_2 s_2 + 2)
                              \right]^{(q_2 -1)/q_2}},
\end{equation}
where $A$ is an integration constant. Therefore at late times when $s_2 \gg 1$, the 
fluctuations will behave like $\delta z \sim (q_2 s_2)^{(2/q_2-1)}$ giving 
$\delta z_r \sim (q_2 s_2)^{-1/2}$ for radiation and 
$\delta z_m \sim (q_2 s_2)^{-1/3}$ for matter demonstrating that the interbrane distance 
will tend to $z_e$ at late times. Incidentally, the early time behaviour 
is $\delta z_r \sim (q_2 s_2)^{-3/4}$ and $\delta z_m \sim (q_2 s_2)^{-2/3}$, and 
although we have already shown that the equilibrium position $z_e$ is not constant 
at this time, provided it moves only slowly relative to the evolution of the 
fluctuations then these early time results will still be of significance. 

\begin{figure}
\center
\epsfig{file=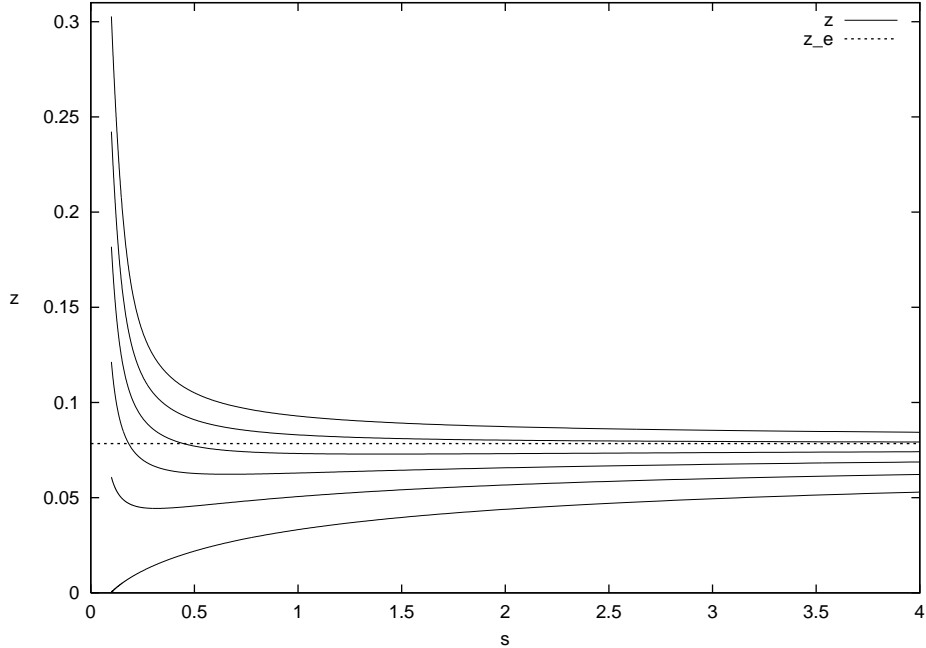,angle=270,scale=0.5}
\caption{The evolution of the interbrane distance $z$ when both branes are 
radiation dominated. The dotted line is the late time value of the 
equilibrium position $z_e$. Note the way some of the trajectories reach a minimum before 
approaching their limit: this is due to the early time dependence of $z_e$}
\label{Three}
\end{figure}

\begin{figure}
\center
\epsfig{file=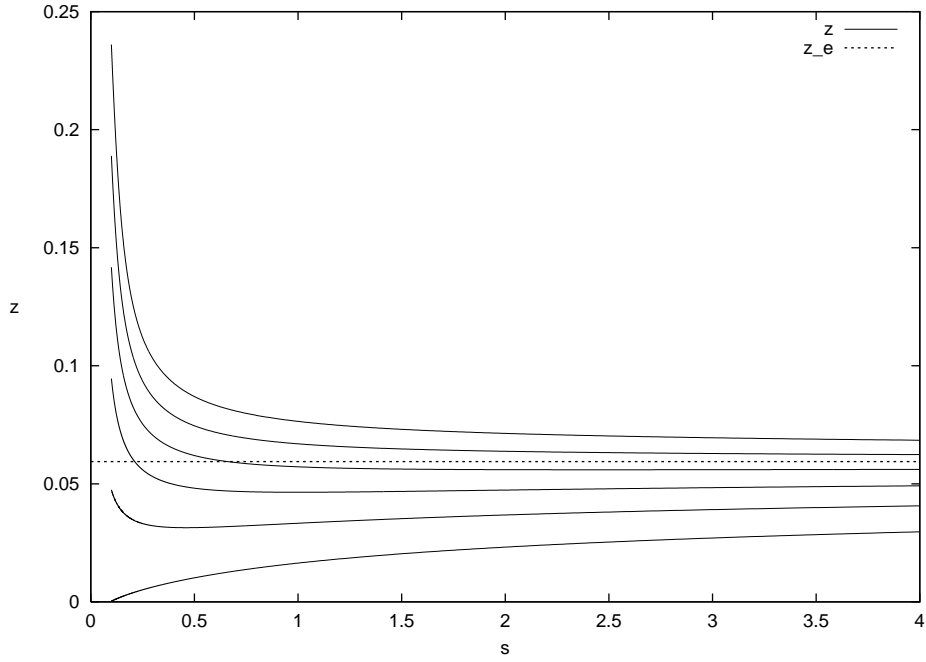,angle=270,scale=0.5}
\caption{Matter on both branes}
\label{Four}
\end{figure}

\begin{figure}
\center
\epsfig{file=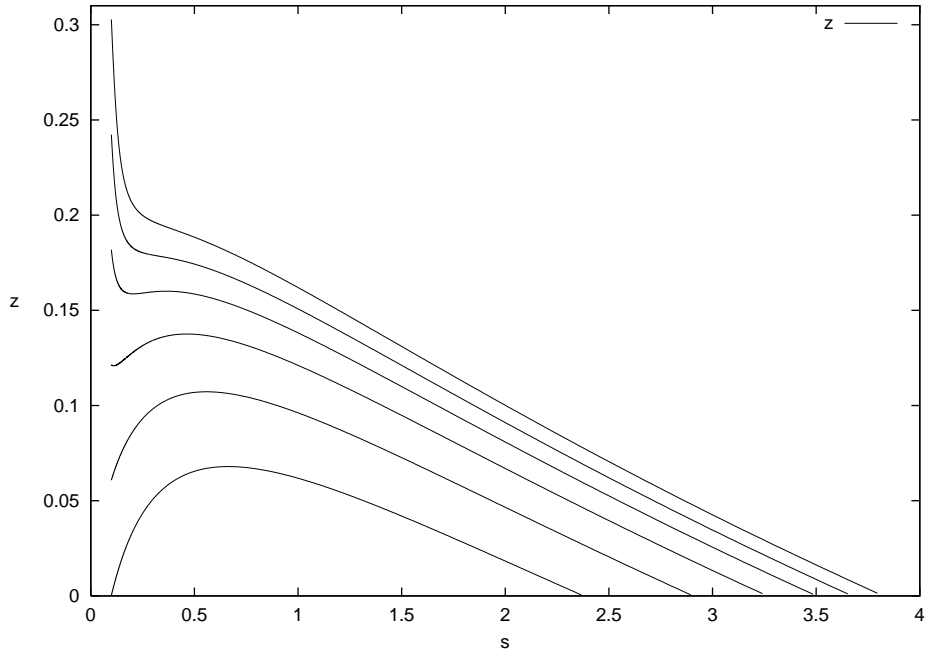,angle=270,scale=0.5}
\caption{Radiation on the reference brane, matter on the second brane}
\label{Five}
\end{figure}

\begin{figure}
\center
\epsfig{file=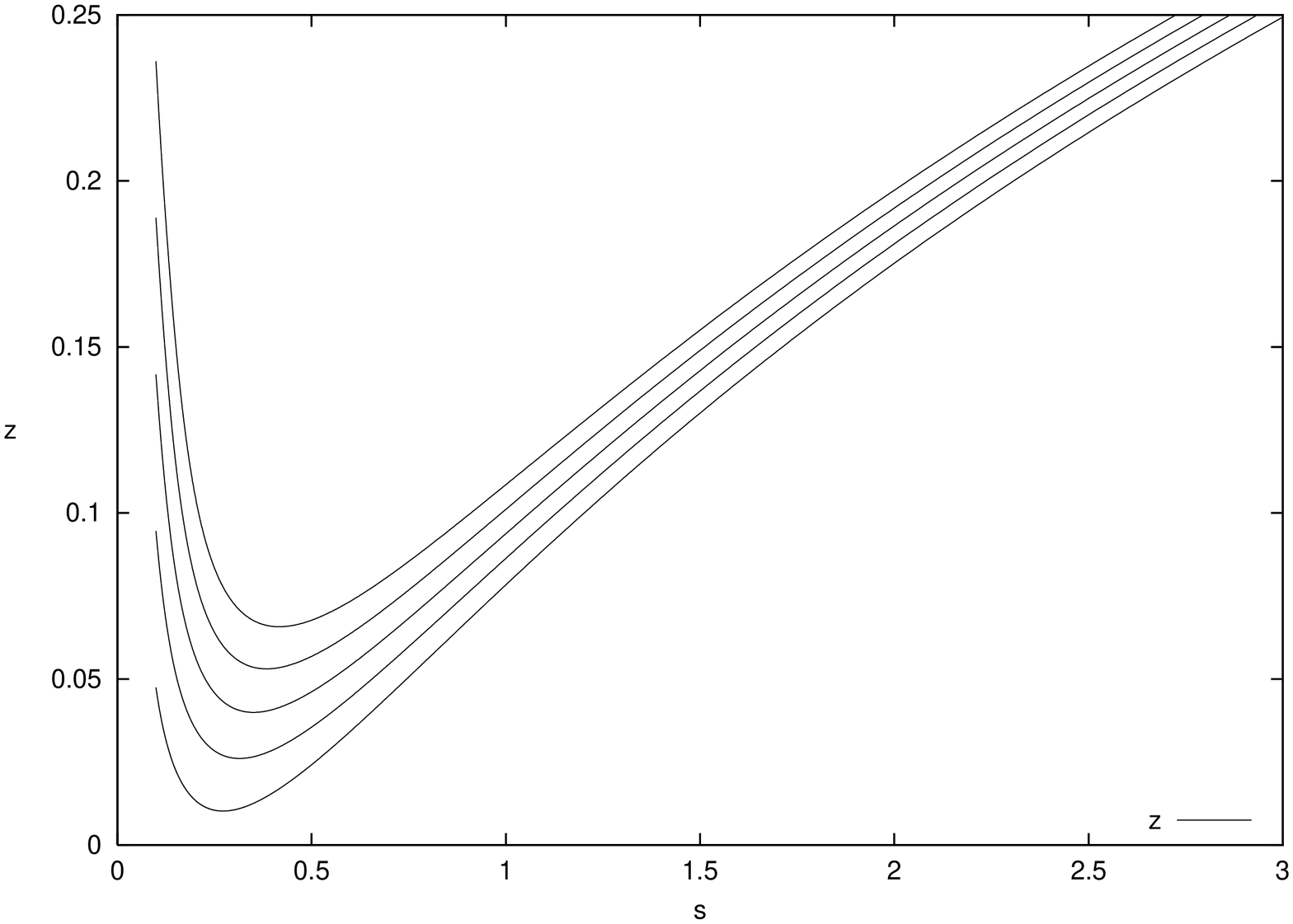,angle=270,scale=0.5}
\caption{Matter on the reference brane, radiation on the second brane}
\label{Six}
\end{figure}

Using equations (\ref{arad}), (\ref{two}), (\ref{assum1}), and (\ref{2ppp}) 
the radion equation (\ref{5R1}) was solved numerically for four different cases, 
with both branes assumed to 
be in a state of either radiation or matter dominance and the results are shown in 
figures~\ref{Three}, \ref{Four}, \ref{Five} and \ref{Six}. In each case the six initial 
radion positions were distributed evenly between $0$ and $z_h$, and the integration 
was started at $s=0.1$. Figure~\ref{Three} shows the evolution of the radion $z$ against 
the dimensionless time variable $s$ when both branes are radiation dominated in that 
$q_0 = q_2 = 4$. The proportionality constants have been chosen to be $\gamma_0 = 1.04$ 
and $\gamma_2 = 1$ and therefore the late time equilibrium position $z_e$ given by 
equation (\ref{zeee}) is $z_e\simeq0.0784$ which is represented by the dotted line. 
At late time the stable behaviour predicted above is evident as all trajectories tend 
slowly towards $z_e$. The expected early time behaviour is also confirmed as initially 
the trajectories rapidly converge, however some of them possess stationary points which 
implies a time dependent solution for $z_e$. In fact, the minimum of a trajectory 
corresponds to where the time dependent solution for $z_e$ actually crosses the trajectory.
Figure~\ref{Four} shows the radion's behaviour when both branes are matter dominated. 
Here the constants are $\gamma_0 = 1.02$ and $\gamma_2 = 1$ and therefore 
$z_e \simeq 0.0594$. A stable behaviour is again observed as expected from the above 
analysis, and the similarities with figure~\ref{Three} are substantial. Note how the 
rate at which trajectories converge at late times is faster for radiation than for 
matter dominated branes in agreement with $\delta z_r \sim s^{-1/2}$ and 
$\delta z_m \sim s^{-1/3}$ as found above. 

The interesting case in which the branes possess different equations of state is 
presented in figures~\ref{Five} and \ref{Six}. Figure~\ref{Five} shows the evolution 
of the radion when the reference brane is radiation dominated and the second brane 
matter dominated. Initially, the trajectories rapidly converge in a similar manner to 
figures~\ref{Three} and \ref{Four}, however soon the disparity in the expansion rates 
of the two branes takes effect and inevitably the interbrane distance reaches zero, 
corresponding to the branes colliding. Figure~\ref{Six} shows the opposite case, 
whereby the reference brane is matter dominated, and the second brane is radiation 
dominated. Again initially the trajectories rapidly converge, but since the reference 
brane is effectively moving faster through the bulk than the second brane, the 
interbrane distance increases indefinitely. 

\subsection{Phase Transitions on the Reference Brane}

In this section we investigate the cosmological behaviour of the 
interbrane distance when phase transitions occur on the reference 
brane. It must be noted that here, unlike the previous section we keep 
the equation of state on the second brane to be $p_2 = - \eta_2$ - that of 
a constant brane tension. 



\begin{figure}
\center
 \epsfig{file=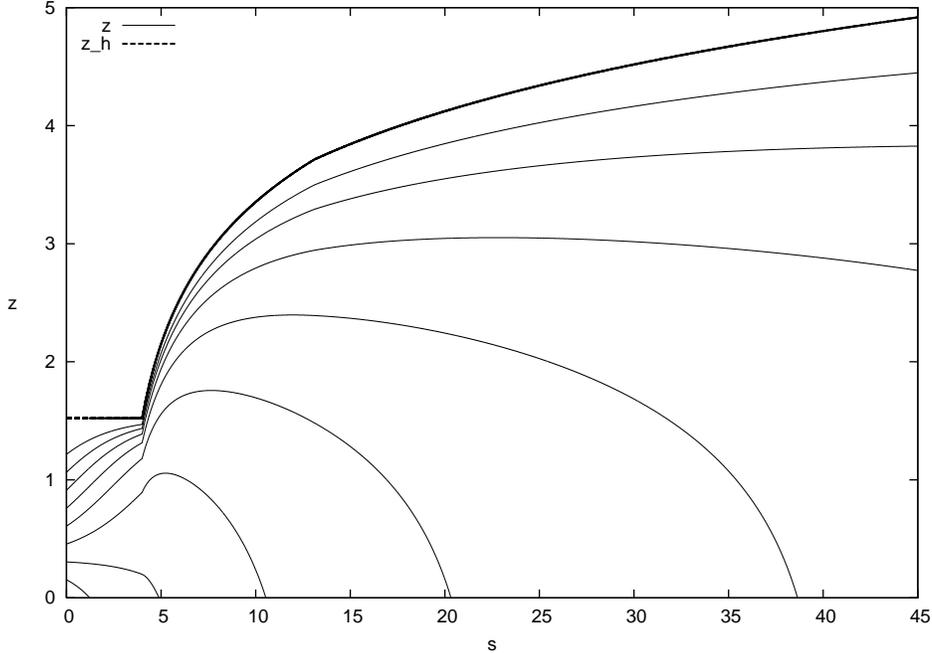,angle=270,scale=0.5}
\caption{Phase transitions: initially the reference brane has a total 
constant energy density $\eta_0 = 1.1$, with $\eta_2 = -1.2$ on
the second brane. At $t=4$ and $t=13$ the reference brane undergoes phase
transitions into a radiation dominated universe and then a matter
dominated universe respectively.}
\label{trans51}
\end{figure}

\begin{figure} 
\center
 \epsfig{file=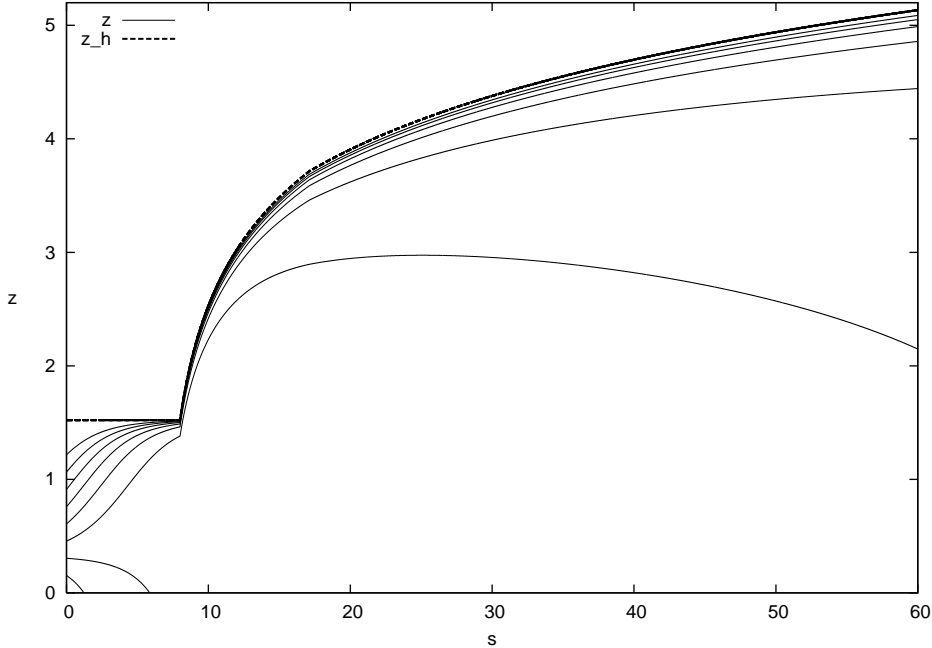,angle=270,scale=0.5}
\caption{The initial phase is twice as long as fig \ref{trans51}
greatly delaying the return of the second brane.}
\label{trans52}
\end{figure}

Here we assume that the reference brane goes through three stages: 
accelerating, radiation dominated and matter dominated. In each stage we 
ignore all subdominant terms in the brane's energy density so that 
$\omega_0 = -1, 0, 1/3$ in each of the corresponding phases, leading to,
\begin{equation}
\eta_{0p} \;\;  = \;\; \left\{ \begin{array}{cc}
\lambda & \;\;  s < s_1 \\
\lambda \frac{a_0(s_1)^4}{a_0(s)^4}  & \;\;  s_1 < s < s_2 \\ 
\lambda \frac{a_0(s_1)^4}{a_0(s_2) a_0(s)^3}  & 
                                               \;\;s_2 < s \\
\end{array}\right. ,
\end{equation}
where $\lambda$ is a constant and $s_1$ and $s_2$ are the times at which 
the transitions occur. Note that here we assume that the transitions
are instantaneous. Using this expression for $\eta_{0p}$ and equations 
(\ref{one}), (\ref{Fri5}), (\ref{arad}) and (\ref{two}) the radion 
equation (\ref{5R1}) was solved for two cases, each with several initial 
radion positions distributed evenly between $z=0$ and $z=0.8z_h$, the 
results of which are 
shown in figures \ref{trans51} and \ref{trans52}. Both cases had 
$\eta_2 = -1.2$ and initially $\eta_0 = 1.1$, and only differed by the 
choice of transition times $s_1$ and $s_2$. The first graph, given in 
figure \ref{trans51} has $s_1=4$, $s_2 \simeq 13$ and shows how during 
the first phase some of the radion 
trajectories quickly collide with the reference brane, and the rest 
tend toward $z_h$ as would be expected from the analysis 
of constant tension branes in section \ref{2AdS}. When the radiation phase 
occurs however, the surviving solutions quickly move away from the 
reference brane (in these brane based coordinates), before eventually 
returning to $z=0$. The existence of the matter phase delays this return 
further. In the second case given by figure \ref{trans52} the initial phase 
is twice as long as in the previous situation with $s_1 = 8$. The 
radiation-matter transition time has been chosen so that matter 
domination occurs at the same energy density as in the first case giving 
$s_2 \simeq 17$. Doubling $s_1$ has an interesting effect - the time taken 
for the inevitable collision between the branes is massively delayed, 
in fact it has been increased roughly by a factor of ten. This shows that 
a relatively short period of stronger inflation on the reference brane 
than on the second brane will delay the return of the second brane 
indefinitely. The fact that in these situations investigated here, the 
interbrane distance eventually reaches zero is of course dependent on the 
value of $\eta_2$ and the limiting value at late time of 
$\eta_0 = 1+\eta_{0p}$. As expected, the behaviour described here and shown 
in the two figures is again in agreement with the conceptual arguments 
in terms of a bulk observer developed in the previous section.

We now go on to examine the interesting case of two positive tension 
non-$Z_2$ symmetric branes in a semi-infinite extra dimension.

\subsection{Non-$Z_2$ Symmetry and Weyl Tensor Effects}

As discussed in section~\ref{MRbored}, it is interesting to examine models where both branes possess 
cosmologically evolving energy densities however, often this requires a second brane 
tension $\eta_2$, such that $\eta_2 \rightarrow -\infty$ as $t \rightarrow 0$. 
This could obviously lead to phenomenological difficulties, hence in this section we 
turn to an interesting model in which the reference brane 
is $Z_2$ symmetric, while the second brane is non-$Z_2$ symmetric allowing 
both branes to be of positive tension in a semi-infinite fifth dimension. 
A model of this kind was previously proposed by~\cite{lykken} as an attempt to 
solve the hierarchy problem using only positive tension branes. Several 
multi-brane extensions of this model have been
studied~\cite{s71,threebrane,s72} albeit not for cosmological cases.

\begin{figure}
 \center
  \epsfig{file=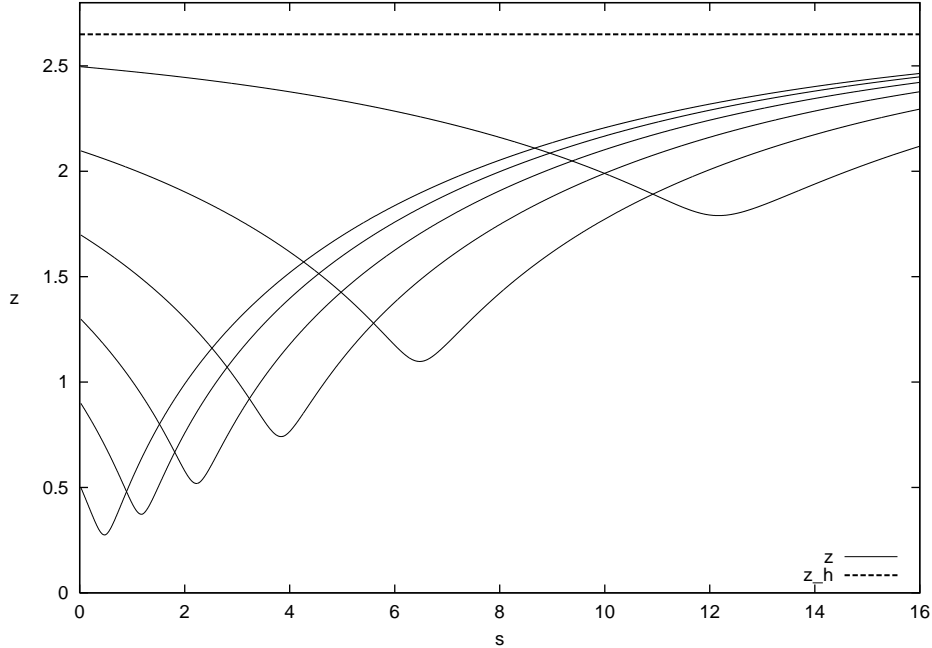,angle=270,scale=0.5}
  \caption{Radion trajectories for the two positive tension branes case where 
           $\mu^3 f_2 = -1$, $\eta_0 = 1.01$ and $\eta_2 = 1.1$}
 \label{fcon}
\end{figure}

Setting $\C_0=\C_1=k=0$ but assuming $\C_2 \not= 0$, means that   
$h_0 = \eta_0^2 - 1$ and that the Friedmann equation for the second brane 
is given by,
\begin{equation}
h_2 \;\; = \;\; 
      \eta_2^2 - 1 \;+\; \frac{\C_2}{2 \mu^2 a_2^4} \;+\; 
        \frac{\C_2^2}{16 \mu^4 \eta_2^2 a_2^8},
\end{equation}
which can be rewritten using $f_2 = \C_2 / 4 \mu $ as,
\begin{equation}\label{fh2}
h_2 \;\; = \;\; 
      \left(
            \eta_2 + \frac{ f_2}{ \mu \eta_2 a_2^4}
      \right)^2 \;-\; 1.
\end{equation}
Referring to equations (\ref{H25}) and (\ref{5R2}) it can be seen that a 
solution of the two brane system with both branes of positive tension 
does exist in this situation, provided that,
\begin{equation}\label{ineq1}
\frac{ - f_2}{ \mu \eta_2 a_2^4} \;\; > \;\; \eta_2.
\end{equation}
This implies both that $f_2<0$ and that the second brane must be 
dominated by its lack of $Z_2$ symmetry, a fact that prevents standard 
cosmology being realised on this brane. We studied the behaviour of the 
interbrane distance for two different equations of state on the second 
brane. The first was where the second brane 
possessed a constant brane tension $\eta_2 = 1.1$ and the results are 
shown in figure \ref{fcon}. The reference brane had a constant tension of 
$\eta_0 = 1.01$ and the non-$Z_2$ symmetric parameter was set to 
$f_2 = -1.0 / \mu^3$ (in order to keep the radion equation independent of
$\mu$) and the six initial radion positions were distributed evenly between 
$z=0.5$ and $z=2.5$. Inspection of equation (\ref{fh2}) shows that in this 
situation the second brane will initially expand until a maximum value of 
$a_2(\tau)$ is reached which satisfies,
\begin{equation}
a_{2max}(\tau) \;\; = \;\;
           \left(
                 \frac{-f_2}{\mu \eta_2 (1+\eta_2)}
          \right)^{1/4}.
\end{equation}
Once this is reached $h_2 = 0 $, and subsequently the second brane begins 
to contract. This behaviour helps to explain the trajectories shown in 
figure \ref{fcon}: the expanding second brane begins to catch up with the 
reference brane but once $a_2 \simeq a_{2max}$, the second brane's motion 
slows and it quickly falls behind and freezes out at $z_h \simeq 2.65$. 

\begin{figure}
 \center
  \epsfig{file=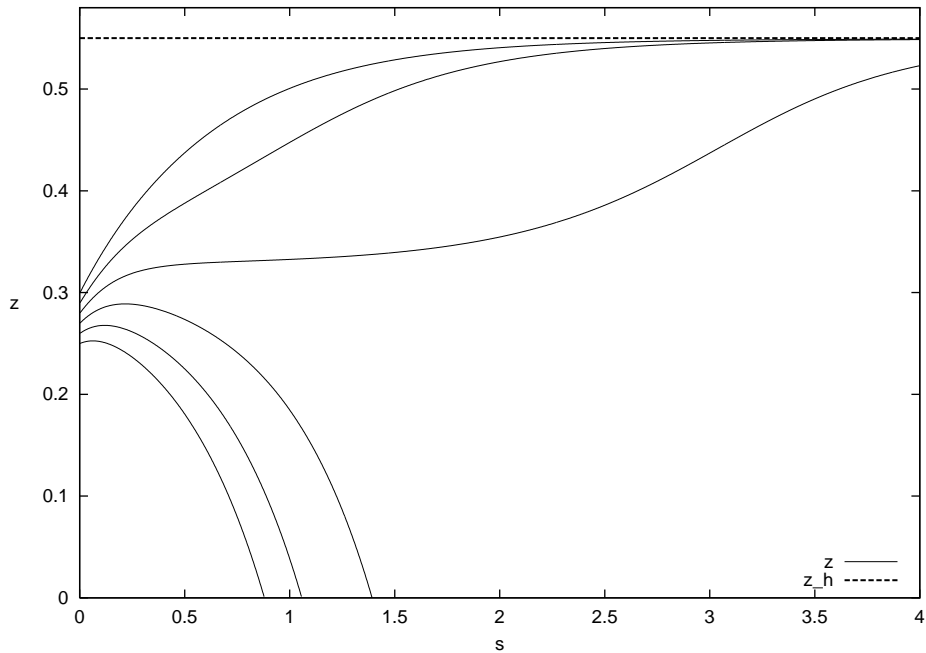,angle=270,scale=0.5}
  \caption{Radion trajectories where the reference brane possesses a 
           constant brane tension and the second brane is radiation 
           dominated and non-$Z_2$ symmetric.}
 \label{frad}
\end{figure}

In the second $f_2 \not= 0$ case investigated, we assumed $\omega_2 = 1/3$ 
and hence that the (positive) energy density $\eta_2$ was radiation 
dominated such that $\eta_2 = \gamma_r/a^4$, where $\gamma_r$ is some 
constant. For each of the six 
trajectories, we set $\gamma_r = 5$ and chose a value of $f_2$ such that 
$\eta_2 + f/\mu \eta_2 a^4 = -1.1$ in order to ensure that the different 
trajectories all start with the same value of $h_2^2=0.21$. The reference 
brane was assumed to have $\eta_0 = 2.0$ and the results are shown in 
figure \ref{frad}. In order to have two positive tension branes the 
inequality (\ref{ineq1}) was maintained and one can therefore see from 
equation (\ref{fh2}) that initially $h_2$ is smaller than $h_0$ and the 
interbrane distance will increase, however the $\eta_2$ term in 
equation (\ref{fh2}) soon becomes negligible and the second brane is then 
entirely dominated by the constant non-$Z_2$ symmetry such that 
$h_2 \simeq f_2^2/ \mu^2 \gamma_r^2 - 1$. Therefore at later times the 
situation is analogous to that of constant tension branes analysed in 
section \ref{2AdS}: an unstable equilibrium position exists as can be 
seen in figure \ref{frad} and this agrees with the perturbative analysis of 
this situation presented in section~\ref{fluc}.  

Obviously there are many other two brane situations that could be 
investigated including both branes having no $Z_2$-symmetry and both 
branes with non-zero Weyl tensor components, however, we leave this to a future 
investigation.

\newpage

\section{Discussion}

In this paper we have greatly extended the work done by Binetruy 
et al~\cite{bin3} on the cosmological radion of the two brane 
Randall Sundrum model. We started by using an elegant method to derive the 
non-linear equations of motion for the interbrane distance $\R$, and hence 
found a more general version of the radion equation, while sidestepping 
a lot of the algebra required in~\cite{bin3}. The $\ddot{\R}$ equation was 
derived and this allowed us to identify the general conditions for an 
equilibrium radion position. The equations of motion were then linearised 
in order to investigate the nature of the equilibrium positions and in the 
majority of cases these positions were found to be unstable, confirming 
and generalising previous work~\cite{radionwave,radionds,locallylocal}.
The radion equations were found to be extremely non-linear and it was 
therefore necessary to numerically integrate them in order to examine the
radion's behaviour far from the equilibrium positions. It should be noted that 
in some cases no equilibrium exists making a numerical analysis even 
more vital. 

The numerical solutions confirmed that for two constant tension $dS$ branes 
there sometimes existed an unstable equilibrium point, but for two $AdS$ 
branes that point, if it exists would be stable. The cases where the reference 
brane possessed a more realistic equation of state, with standard cosmology 
taking over at late times were also examined, and for a
constant second brane tension $\eta_2>1$ it was found that the branes 
inevitably collide. The time taken for this collision to occur was calculated 
and shown to depend upon amoung other things the initial radion position and 
the nature of the matter on the reference brane, for example, 
radiation domination implies the branes collide sooner than if there was 
matter domination. It was demonstrated how if both branes possess a time 
dependent energy density then a stable equilibrium solution exists provided 
the branes possess the same equation of state. The effects of changes of state 
were also considered, and it was 
shown how a relatively short burst of inflation could greatly delay the 
return of the second brane. Two non-$Z_2$ symmetric scenarios involving two 
positive tension branes were investigated: one led to all trajectories 
freezing out at $z_h$, while the other at late times was analogous to 
the the constant tension $dS$ case. It should be noted that these scenarios 
exist in a semi-infinite space $y>0$, and assuming 
$f_2<0$ implies that the Schwarzschild mass to the right of the second brane 
$\C_2<0$ which may lead to problems as the second brane 
will not be shielded from the naked singularity appearing in the bulk. This 
could be avoided by having for example $\C_0=\C_1>\C_2>0$ i.e. having all 
the Schwarzschild masses greater than zero. 

It was discussed how, by considering the perspective of a bulk observer in 
a static bulk background, one can develop a conceptual understanding 
of many of the features of the radion displayed in this paper. Viewing 
the behaviour of the interbrane distance as a competition between the 
expansion rates of the branes, one can then analyse the branes' Friedmann
equations to determine qualitatively the radion's behaviour. One may ask 
whether it would have been better to carry out this analysis in the bulk 
based coordinate system as opposed to a brane based one. However, we intend 
to extend this radion analysis to systems containing more complicated bulk 
matter than just a cosmological constant, namely scalar and dilaton fields. 
It is then not possible to use a time independent bulk based coordinate 
system since Birkhoff's theorem (the generalisation of Gauss' theorem) can 
no longer be applied. Given the elegance and ease 
of deriving the radion equation in the brane based system, the need to 
follow and extend the work done by~\cite{bin3}, and the inevitable 
numerical calculation whatever coordinate system was used, it was 
felt that the brane based system was the most appropriate for our 
purposes.

Finally, it should be mentioned how special a case the cosmological radion 
is. Due to the amount of symmetry assumed, the trajectories of the branes are 
entirely determined by the knowledge of their Friedmann equations and hence 
their expansion rate, a fact which allowed us to efficiently derive the 
radion equation, and was not realised by~\cite{bin3} who had to use a more 
cumbersome method. In essence the branes do not `feel' each other, a fact 
that is altered if one considers other fields in the bulk, or general 
perturbations to the cosmological background.

\vspace{0.5cm}\noindent {\bf Acknowledgements:}
We would like to thank Pierre Binetruy and Carsten van de Bruck for useful discussions. 
This work is supported in part by PPARC.

%
\bibliographystyle{JHEP} \bibliography{thesis1}
%

%
\end{document}